\newif\ifproofs
\proofstrue 
\newif\ifjournalappendix
\journalappendixtrue
\newif\ifcorroltwo
\corroltwotrue

\documentclass[10pt,final,twocolumn]{IEEEtran}
\IEEEoverridecommandlockouts

 \usepackage{beb_tikzstyle}
 \usepackage{bebstyle}

\usepackage{enumitem,color,amsthm}

\usepackage[ruled,linesnumbered,algo2e]{algorithm2e}  
\SetKwInput{KwData}{\textbf{Input}} 
\SetArgSty{textnormal}
\let\oldnl\nl
\newcommand{\nonl}{\renewcommand{\nl}{\let\nl\oldnl}}
\usepackage{algpseudocode}
\usepackage{cleveref}
\crefrangelabelformat{section}{#3#1#4--#5\crefstripprefix{#1}{#2}#6}
\crefformat{equation}{(#2#1#3)}
\Crefname{figure}{Fig.}{Figs.}
\crefrangeformat{equation}{eqs. #3(#1)#4--#5(#2)#6}

\newtheorem{theorem}{Theorem}
\newtheorem{corollary}{Corollary}
\newtheorem{lemma}{Lemma}

\makeatletter
\newcommand*{\org@overidelabel}{}
\let\org@overridelabel\@verridelabel
\@ifpackagelater{acronym}{2015/03/21}{
  \renewcommand*{\@verridelabel}[1]{%
    \@bsphack
    \protected@write\@auxout{}{\string\AC@undonewlabel{#1@cref}}%
    \org@overridelabel{#1}%
    \@esphack
  }%
}{
  \renewcommand*{\@verridelabel}[1]{%
    \@bsphack
    \protected@write\@auxout{}{\string\undonewlabel{#1@cref}}%
    \org@overridelabel{#1}%
    \@esphack
  }%
}
\makeatother

\renewcommand{\baselinestretch}{0.99}

\begin{document}

\title{Anomaly Search over Composite Hypotheses in Hierarchical Statistical Models
%
}

\author{
\IEEEauthorblockN{Benjamin Wolff, Tomer Gafni, Guy Revach, Nir Shlezinger, and Kobi Cohen (\emph{Senior Member, IEEE})}
\thanks{B. Wolff and T. Gafni contributed equally to this work.}
    \thanks{
A short version of this paper that introduces the algorithm, and preliminary simulation results was presented at the IEEE International
Symposium on Information Theory (ISIT) 2022 \cite{Wolff2022hierarchical}. In this journal version we include: (i) A more general model that includes multiple anomalies; (ii) a detailed discussion on the implementation of
the algorithm; (iii) a rigorous theoretical analysis of the algorithm with detailed proofs; (iv) more extensive simulation results including new experiments using synthetic and real data; and (v) a detailed discussion of the results, and comprehensive discussion and comparison with the existing literature.

B. Wolff and G. Revach are with the Institute for Signal and Information Processing (ISI), D-ITET, ETH Zürich, Switzerland, 
		(e-mail: bewolff@student.ethz.ch; grevach@ethz.ch).
		T. Gafni, N. Shlezinger, and K. Cohen are with the School of Electrical and Computer Engineering, Ben-Gurion University of the Negev, Beer-Sheva, Israel (e-mail:gafnito@post.bgu.ac.il; \{nirshl, yakovsec\}@bgu.ac.il).
This research was partially supported by the ISRAEL SCIENCE FOUNDATION
(grant No. 2640/20), and by the Israel National Cyber Directorate via the Cyber Security Research Center at Ben-Gurion University of the Negev.
}
\vspace{-1.0cm}
}

\maketitle
\pagenumbering{arabic}
\begin{abstract}
Detection of anomalies among a large number of processes is a fundamental task that has been studied in multiple research areas, with diverse applications spanning from spectrum access to cyber-security. 
Anomalous events are characterized by deviations in data distributions, and thus can be inferred from noisy observations based on statistical methods. In some scenarios, one can often obtain noisy observations aggregated from a chosen subset of processes. Such hierarchical search can further minimize the sample complexity while retaining accuracy. An anomaly search strategy should thus be designed based on multiple requirements, such as maximizing the detection accuracy; efficiency, be efficient in terms of sample complexity; and be able to cope with statistical models that are known only up to some missing parameters (i.e., composite hypotheses).
In this paper, we consider anomaly detection with observations taken from a chosen subset of processes that conforms to a predetermined tree structure with partially known statistical model. We propose \ac{hds}, a sequential search strategy that uses two variations of the \ac{gllr} statistic, and can be used for detection of multiple anomalies. \ac{hds} is shown to be order-optimal in terms of the size of the search space, and asymptotically optimal in terms of detection accuracy. An explicit upper bound on the error probability is established for the finite sample regime. In addition to extensive experiments on synthetic datasets, experiments have been conducted on the DARPA intrusion detection dataset, showing that \ac{hds} is superior to existing methods.
\end{abstract}


\acresetall 
\section{Introduction}\label{sec:Introduction}
The task of detecting anomalies in data streams arises in a wide variety of applications. These applications include dynamic spectrum access and sensing in wireless communication \cite{zhao2007survey}; detecting attacks and intrusions in computer networks \cite{zhang2006anomaly}; and 
detecting anomalies in infrastructures that may indicate catastrophes \cite{genge2014connection}. 
Such tasks involve distinguishing anomalous processes from typical ones based on noisy observations. 

The noisy nature of the observations implies that the typical behavior can be modeled by a normal or benign distribution, and the anomalous behavior is captured by an abnormal distribution. The goal of a \acl{dm} boils down to deciding whether to reject the null hypothesis and to declare a process as anomalous. Here we consider the task of detecting an anomalous process (or processes) out of a large set of data streams. This requires to sample (observe) each process at least once,  and preferably more for better detection accuracy due to uncertainty. Hence, a \acl{dm} should design an \emph{efficient} sampling policy, that for a given detection accuracy minimizes the number of samples needed to reach a decision, or alternatively, given a sampling budget maximizes the detection accuracy.

The class of problems involving a sequential design of experiments for active binary hypothesis testing problem was pioneered by Chernoff \cite{chernoff1959sequential}. Chernoff proposed a randomized strategy and showed that it is asymptotically optimal as the error probability approaches zero. However, the Chernoff test results in a linear sample complexity in the size of the search space.  When the number of processes (data streams) is very large, as is often the case in practice, it is likely to be inefficient to sample each process multiple times. Therefore, sampling strategies with a sub-linear sampling complexity are desirable. 
The need for efficiency requires to exploit a certain structure in the data, which may lead to a significant performance gain. A common structure that can be utilized for this end is the ability to access the data in a hierarchical fashion. 

The hierarchical structure model represents settings where a massive number of data streams can be observed at different levels of granularity. Such modeling faithfully captures the operation of various applications. In finance, transactions can be aggregated at different temporal and geographic scales \cite{ahmed2016survey}. In visual monitoring applications, the ability to zoom-in or zoom-out is equivalent to the aggregation of pixels, and can lead to faster detection of anomalies (targets, interesting events) by giving suspicious pixels more attention than others \cite{singh2020crowd}. In internet traffic monitoring, there is a need for detecting heavy hitters, i.e., a small number of flows that accounts for most of the total traffic, and thus representing the measurements as a tree structure, where each node represents an aggregated flow can lead to  efficient detection \cite{thompson1997wide}. Other applications include direction of arrival estimation \cite{chiu2019active} and system control~\cite{simsek2004scalar}.

In light of the aforementioned potential gains of the hierarchical structure, here, we consider the problem of detecting anomalous processes (targets), for which there is uncertainty in the distribution of observations. We assume that in each time step, a decision-maker can observe a chosen subset of processes that conforms to a predetermined tree structure, and get access to aggregated observations that are drawn from a general distribution that depends on a chosen subset of processes. The uncertainty in the anomalous distribution yields  a composite hypothesis case, where measurements drawn when observing a subset of processes follow a common distribution parameterized by an unknown vector when containing the target.  The objective is to design a sampling policy (a search strategy), that minimizes a Bayesian risk that accounts for sample complexity and detection accuracy, by selecting which subset to observe, and when to terminate the search and make a decision, in an adaptive way. 

%
Dynamic search strategies were proposed for various forms of anomaly detection problems. 
In \cite{wang2020information}, the \ac{irw} algorithm was proposed, for cases where the statistical model is fully known. \ac{irw} was  shown to be asymptotically optimal in terms of detection accuracy and order optimal with respect to the number of processes. When the anomalous hypothesis is composite, the \ac{irw} policy serves as a benchmark for the performance one can achieve with partially known modeling. 
The recent studies \cite{vakili2018hierarchical,vakili2019random,gafni2021searching} considered hierarchical search under unknown observation models. The search strategies in \cite{vakili2018hierarchical,vakili2019random} are based on a sample mean statistic, which fails to detect a general anomalous distribution with a mean close to the mean of the normal distribution. The work in \cite{gafni2021searching} does not assume a structure on the abnormal distribution, and uses the Kolmogorov-Smirnov statistic, which fails to utilize the parametric information considered in our setting. This motivates the derivation of a dynamic search policy for data of hierarchical structure which can cope with partially known anomalous distributions and reliably detect based on statistics of a higher order than a sample mean.

%
In this work we consider for the first time the task of hierarchical anomaly detection over a general and known family of distributions with unknown parameters. Here, the measurements can take continuous values and the decision-maker is allowed to sample an aggregated subset of processes that conforms to a tree structure. To cope with this observation model in a dynamic search setting with possibly multiple anomalies of different types, we develop a dedicated sequential search strategy, dubbed \ac{hds}. \ac{hds} uses two carefully chosen statistics to harness the information on the  null hypothesis and the structure of the hierarchical samples, allowing it to achieve asymptotically optimal performance. The proposed policy is shown to be asymptotically optimal with respect to the detection accuracy and order optimal with respect to the size of the search space. 

Extensive numerical experiments on synthetic and real datasets support the theoretical results. Our numerical evaluation shows that \ac{hds} effectively captures changes in the traffic that are associated with network anomalies. \ac{hds} with active local tests for the high level nodes is also analyzed numerically and is shown to outperform the fixed sample-size local test and approach the performance bound of \ac{irw}. 
Our non-synthetic experiments numerically evaluate the performance of \ac{hds} in a cyber-security  task using the DARPA intrusion detection dataset. We show that the proper modeling of the network traffic data in a hierarchical fashion combined with the application of \ac{hds} can successfully detect \ac{dos} attacks from a limited number of samples. 

The rest of this paper is organized as follows: in Section \ref{sec:System} we present the system model and discuss its relationship with the existing literature. Section \ref{sec:HDS} designs the \ac{hds} policy and analyzes its performance. We numerically evaluate \ac{hds} in Section \ref{sec:Simulations}, and provide concluding remarks in Section \ref{sec:Conclusions}.

\section{System Model and Preliminaries}\label{sec:System}
In this section, we describe the statistical setting of our system model and discuss some of the relevant related literature on dynamic search policies.
%
%
\subsection{Problem Formulation}\label{subsec:Problem}
{\bf Anomaly Detection:} 
We consider the problem of detecting $\nanom$ anomalous (targets) processes (data streams) out of a large set of $\nproc$ processes, where $\nanom$ is assumed to be known. Here, the decision-maker should first actively collect evidence (data, observations, samples), and then decide for each process $\iproc\in\set{1,..,\nproc}$ whether it is anomalous or not. Since there is cost on gaining samples, this problem presents an inherent trade-off between the need to maximize the detection accuracy to the need to minimize the length of the exploration phase.  

In particular, in each time step $t$, where $t\in\set{1,2,\ldots}$, the decision-maker can access only one process and sample an observation $\obs_t$ in an \acs{iid} manner. The main challenge is to know when to stop exploring and to reach a decision. We denote the data collected in the time horizon $\tau$ and provide a decision based on $\mathbf{D}_\tau=\set{\obs_{t}}_{t=1}^\tau$. Given the collected evidence, the decision rule boils down to  \emph{simultaneous} testing of multiple binary hypotheses. Let $\hypo_{\iproc}=0$ denote the \emph{null hypothesis}, i.e., the process $\iproc$ is not anomalous, then the decision-maker should decide whether to reject the null hypothesis, and declare process $\iproc$ as anomalous, i.e., $\hypo_{\iproc}=1$, or not. 

Assuming that at time $t$, process $m$ was chosen to be sampled by the decision-maker, then its sampling distribution is given by
\begin{equation}\label{eq:ref_flat_model}
\obs_t\sim\density\brackets{y\gn\paramvec},
\quad
\left\{ 
  \begin{array}{ c l }
    \paramvec=\paramvec_0, & \quad \hypo_{\iproc}=0\\
    \paramvec\in\paramset_1, & \quad \hypo_{\iproc}=1
  \end{array}
\right.
\end{equation}
where $f\brackets{\cdot}$ is a known family of a parametric probability distributions. While $\paramvec_0$ is a known parameter describing the distribution of the non-anomalous samples, for anomalous processes, the parameter is not assumed to be known, but only that it is restricted to belong to a known set $\paramset_1$.


{\bf Hierarchical Sampling:} 
To reach a decision, the decision-maker must actively sample information from the $\nproc$ processes. Generally speaking, if the complexity of an active sampling policy is linear, when the number of processes $\nproc$ scales up, such policy becomes inefficient and can be computationally infeasible. Therefore, to cope more efficiently with a large number of processes, and to reduce the sampling complexity,  we consider the case of hierarchical data streams. Here, in addition to observing individual processes, the decision-maker can measure aggregated processes that conform to a binary tree structure. Sampling an internal node of the tree gives a blurry image of the processes beneath it, as schematically depicted in Fig.~\ref{fig:binary_tree}. The key to utilizing the hierarchical structure of the sampling space to its full extent, is to determine the number of samples one should obtain at each level of the tree, and when to zoom in or out on the hierarchy. 
%

To model hierarchical sampling, let the tuple $(l,j)$ denote node $j$ at level $l$ of the tree, with $l=0,\ldots,\levels$ and  $j=1,\ldots,2^{\levels-l}$.
The tree structure encodes the relationship between the nodes. The abnormal distribution of a target leads to an abnormal distribution in every ancestor of the target, i.e., every node on the shortest path from this target to the root.
We denote by $\hypo_{(l,j)}=0$ the hypothesis that node $(l,j)$ is not anomalous, and $\hypo_{(l,j)}=1$ denotes that it is anomalous.

The observations $\obs_t$ of an internal node $j$ on level $\level$ of the tree follow a similar statistical model as in \eqref{eq:ref_flat_model}: %
\begin{equation}\label{eq:ref_hi_model}
\obs_t\sim\density_\level\brackets{y\gn\paramvec},
\quad
\left\{ 
  \begin{array}{ c l }
    \paramvec=\paramvec_0^{\brackets{\level}}, & \quad \hypo_{(l,j)}=0\\
    \paramvec\in\paramset_1^{\brackets{\level}}, & \quad \hypo_{(l,j)}=1
  \end{array}
\right.
\end{equation}
where $\density_\level \brackets{\cdot}$, $\paramvec_0^{\brackets{\level}}$, and $\paramset_1^{\brackets{\level}}$ are the probability distribution, the known parameter of the non-anomalous distribution, and the set of the anomalous parameter, respectively, at level $\level$, and $\density_0 \brackets{\cdot} \equiv \density \brackets{\cdot}$, $\paramvec_0^{\brackets{0}} \equiv \paramvec_0$, $\paramset_1^{\brackets{0}} \equiv \paramset_1$. 
Here, we assume that the observations at all levels are informative, as formulated in the following:
	\begin{enumerate}[label={\em AS\arabic*}, resume]   
	\item \label{itm:distinguish} The \ac{kl} divergence $\div_{\level}\brackets{\cdot||\cdot}$ between  $f_{\level}\brackets{\cdot \gn x}$ and $f_{\level}\brackets{\cdot\gn z}$ satisfies: 
	\begin{equation}\label{eq:assumption}
	\div_{\level}\br{\paramvec_0^{(\level)}||\paramvec}\geq\mindelta, \quad \div_{\level}\br{\paramvec||\paramvec_0^{(\level)}}\geq\mindelta,\quad\forall\paramvec \in\paramset_1^{(\level)},
	\end{equation}
	for some $\mindelta>0$ independent of $M$ for all $l$.
\end{enumerate}
Note that \ref{itm:distinguish} implies that the anomalous and non-anomalous distributions are distinguishable.

{\bf Search Policy:} 
An active search (i.e., sampling and decision) policy (strategy), denoted as $\policy$, is defined by the tuple $\br{\select,\stop,\dec}$. Here, $\select$ is a sampling selection rule, i.e., a mapping from the time $t$ and the data collected to a node from which we need to sample next, namely
\begin{equation}
\select:\brackets{t,\mathbf{D}_t}\mapsto (\level,j);
\end{equation}
a stopping rule $\stop$, i.e., the time at which the decision-maker decides to end the search; and a decision rule $\dec$ 
\begin{equation}
\dec:\brackets{\mathbf{D}_\tau}\mapsto\left[0,1\right]^\nproc,
\end{equation}
which is a mapping from the evidence collected until stopping time $\tau$, to a Boolean vector of size $\nproc$, where $\dec_m=1$ corresponds to the decision that process $\iproc$ is anomalous.

{\bf Aim:} We aim  to find a policy $\policy^\ast$ out of the set of possible policies $\Policy$ that minimizes the \emph{Bayesian risk}, namely
\begin{equation}
\policy^\ast=\arg\min_{\policy\in\Policy}\set{\risk\br{\policy}},
\end{equation}
where
\begin{equation}
\label{eqn:Risk}
\risk\br{\policy}\triangleq\mathcal{P}_{\mathrm{Err}}\brackets{\policy}+\cost\cdot\complexity\br{\policy}. 
\end{equation}
The term $\er\brackets{\policy}$ is the error probability, 
$\complexity\br{\policy}$ is the sample complexity, and $c \in (0,1)$ is the sampling cost assigned for each observation. 
%
%
Specifically, let $\mathcal{H}$ be a set of all Boolean vectors of size $\nproc$ with exactly $K$ entries equal to $1$, such that $|\mathcal{H}|= {M \choose K}$. Let $H_b\in\mathcal{H}$ be any Boolean vector of size $\nproc$ with exactly $K$ entries equal to $1$, and $H\in\mathcal{H}$ be a Boolean vector of size $\nproc$ corresponding to the true hypothesis, where an $m$th entry equals to $1$ implies that process $m$ is anomalous. Then, the error probability given that $H=H_b$ is:
\begin{equation}\label{eqn:errorRate_single}
\mathcal{P}_{\mathrm{Err}}\brackets{\policy|H=H_b}\triangleq
\mathcal{P}\brackets{\dec\brackets{\mathbf{D}_\tau}\neq H_b | \policy, H=H_b},
\end{equation}
and the error probability is averaged under given prior $p_b$ for hypothesis $H=H_b$:
\begin{equation}\label{eqn:errorRate}
\mathcal{P}_{\mathrm{Err}}\brackets{\policy}\triangleq
\sum_{H_b\in \mathcal{H}}
p_b\cdot \mathcal{P}_{\mathrm{Err}}\brackets{\policy|H=H_b}.
\end{equation}
Similarly, the sampling complexity is averaged under prior $p_b$ for hypothesis $H=H_b$, and is given by:
\begin{equation}\label{eqn:complexity}
\complexity\br{\policy}\triangleq\expect \sbr{\stop| \policy}.
\end{equation}
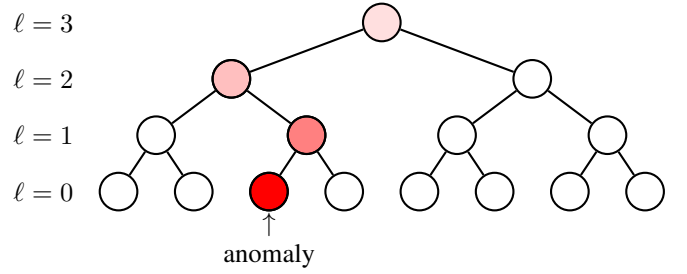
\begin{figure}[tbp]
\centerline{
\begin{tikzpicture}[
    every path/.append style={thick},
    scale=.5
    ]
    \def\off{1}
    \def\offs{1.3}
    \def\yoff{.7}
    \node[node2, fill=red] (anom) at (-3,0) {};
    \node[node2, fill=red!50] (anom) at (-2,1.5) {};
    \node[node2, fill=red!25] (anom) at (-4,3) {};
    \node[node2, fill=red!12.5] (0) at (0,4.5){};
    \foreach\x [count=\xi] in {-4,4}{
        \node[node2] (\x) at (\x,3){};
        \draw (\x) -- (0);
        \foreach\y [count=\yi] in {\x-2,\x+2}{
            \node[node2] (\y) at (\y,1.5){};
            \draw (\x) -- (\y);
            \foreach\z [count=\zi] in {\y-1,\y+1}{
            \node[node2] (\z) at (\z,0){};
            \draw (\z) -- (\y);
            \node at (\y+\off,0-1){};
        }
        }
    }
    \node (anomtext) at (-3,-1.3) [align=center] {$\uparrow$\\anomaly};
    \foreach\h [count=\l] in {0,1.5,3,4.5}{
        \pgfmathsetmacro\ldef{\l-1}
        \node (level) at (-9,\h) {$\level=\pgfmathprintnumber{\ldef}$};
    }
\end{tikzpicture}
}
\vspace{-0.2cm}
\caption{A binary tree observation model with $\nproc=8$ processes, $\levels=3$ levels, and a single anomaly. The anomaly is measurable at the red nodes.}
\vspace{-0.2cm}
\label{fig:binary_tree}
\end{figure}
%
%
\subsection{Related Literature}\label{ssec:review}
Target search problems have been widely studied under various scenarios.
Optimal policies for target search with a fixed sample size were derived in \cite{tognetti1968optimal,kadane1971optimal,zhai2013dynamic,castanon1995optimal} under restricted settings involving  binary measurements and symmetry assumptions.
Results under the sequential setting can be found in \cite{ zigangirov1966problem,klimko1975optimal},  
assuming single process observations. In this paper we address the optimality question under the asymptotic regime as the error probability approaches zero. Asymptotically optimal results for  sequential anomaly detection in a linear (i.e., non-hierarchical) search under various setting can be found in \cite{cohen2014optimal,huang2018active,gurevich2019sequential,lambez2021anomaly}.
In this paper, however, we consider a composite hypothesis case, which was not addressed in the above. Results under the composite hypothesis case with linear search can be found in \cite{vaidhiyan2017learning,nitinawarat2015universal,cohen2015asymptotically,tartakovsky2014nearly,tartakovsky2020nearly}. 
Detecting anomalies or outlying sequences has also been studied under different formulations, assumptions, and objectives \cite{caromi2012fast, heydari2016quickest, geng2016quickest, tajer2013quick, tsopelakos2019sequential, tsopelakos2022sequential}; see survey in \cite{tajer2014outlying}. These studies, in general, do not address the optimal scaling in the detection accuracy or the size of the search space.

As mentioned in Section \ref{sec:Introduction}, the problem considered here also falls into the general class of sequential design of experiments pioneered by Chernoff in  \cite{chernoff1959sequential}. Compared with the classical sequential hypothesis testing pioneered by Wald \cite{wald2004sequential} where the observation model under each hypothesis is fixed, active hypothesis testing has a control aspect that allows the decision-maker to choose different experiments (associated with different observation models) at each time.
The work \cite{naghshvar2013active}  developed a variation of Chernoff’s randomized test that achieves the optimal logarithmic order of the sample complexity in the number of hypotheses under certain implicit assumptions on the \ac{kl} divergence between the observation distributions under different hypotheses. These assumptions, however, do not always hold for general observation models as considered here.

In contrast to Chernoff's randomized policy, in this paper we propose an active \textit{deterministic} strategy. The work \cite{cohen2015active} have showed that a simpler deterministic algorithm applies in this setting and obtained the same asymptotic performance as Chernoff's policy, with better performance in the finite sample regime under a linear search setting with known distributions. A modified algorithm has been developed in \cite{egan2017fast} for spectrum scanning with time constraint.
This setting was extended in \cite{hemo2020searching} to the composite case, which proposed an asymptotically optimal deterministic policy. The problem addressed in this work is fundamentally different, focusing on efficient exploitation of aggregated and potentially low-quality measurements to achieve an optimal sublinear order with the size of the search space. 

Tree-based search in data structures is a classical problem in computer science (see, for example, \cite{bentley1975multidimensional,sleator1985self}). It is mostly studied in a deterministic setting; i.e., the observations are deterministic when the target location is fixed. The problem studied in this work is a statistical inference problem, where the observations taken from the tree nodes follow general statistical distributions.  
This problem also has intrinsic connections with several problems studied in different application domains.
We discuss here two representative studies most pertinent to this paper and emphasize the differences in our approach from these existing studies:\\
$1)$ The first is  group testing, where the objective is to identify the defective items in a large population by performing tests on subsets of items that reveal whether the tested group contains any defective items. Formulations of group testing can be mapped to our setting by mapping the individual items to the leaf nodes of a tree. The action of testing a node on the tree corresponds to a group test. 
Differ from our setting, most existing work on Boolean group testing assumes error-free test outcomes, or limited noise models (e.g., binary symmetric noise or one-sided noise \cite{atia2012boolean,tan2014strong}). Moreover, most of the existing results on noisy group testing focus on non-adaptive open-loop strategies \cite{kaspi2015searching,scarlett2018noisy}, and the issue of sample complexity in terms of the detection accuracy is absent in the basic formulation.
\\
$2)$ Our setting also applies to adaptive sampling with noisy response, for example, in the fundamental problem of estimating a step function in $[0,1]$ \cite{frazier2019probabilistic}. The main body of work on adaptive sampling is based on a Bayesian approach with binary noise of a known model.
Although several strategies (e.g., the Probabilistic Bisection Algorithm) have been extensively studied in the literature \cite{waeber2013bisection,ben2008bayesian}, there is little known about the theoretical guarantees, especially when it comes to unknown noise models. HDS, derived in the sequel based on the problem formulated in Subsection~\ref{subsec:Problem} can be considered as a non-Bayesian approach to the adaptive sampling problem under general parametric noise models, and its theoretical guarantees apply in this problem.

\color{black}
%
%
\section{Hierarchical Dynamic Search}\label{sec:HDS}
In this section we present and analyze the propose\textcolor{NewColor}{d} \ac{hds} active search strategy. We start by introducing the algorithm in the case of one anomaly (i.e., $K=1$) in Subsection~\ref{subsec:HDSDesign}, after which we analyze its performance in Subsection~\ref{subsec:analysis}. In Subsection~\ref{subsec:multi} we extend HDS to multi-target setting, and we conlclude the section with a discussion in Subsection~\ref{subsec:discussion}.
%
%
%

%
%
\subsection{Algorithm Design}\label{subsec:HDSDesign}
We start by focusing on detecting a single target ($K=1$).

{\bf Rationale:} The anomaly is searched using a random walk on the process tree that starts at the root node. The individual steps of the walk are determined by local tests.
On internal (i.e., high level) nodes, the outcome of the test can be moving to the left or right child, or returning to the parent node (where the parent of the root  is itself).
The internal test is constructed to create a bias in the walk towards the anomalous leaf.
On a leaf node of index $\iproc$, the possible outcomes are either terminating the search and declaring process $\iproc$ anomalous, or moving back to parent node. The leaf test is designed to terminate at the  anomaly with sufficiently high probability.

In particular, \ac{hds} uses the fixed sample size \ac{gllr} statistic for the high level nodes test and the sequential \ac{allr} statistic for the leaf nodes test.
The \ac{allr} statistic, introduced by Robbins and Siegmund \cite{robbins1972class,robbins1974expected}, builds upon the one-stage delayed estimator of the unknown parameter; i.e., the density of the $n$-th observation is estimated based on the previous $n - 1$ observations, while the current  observation is not included in this estimate. As opposed to the \ac{gllr}, the \ac{allr} preserves the martingale properties. This allows one to choose thresholds in a way to control specified rates of error probability, and so to ensure the desired asymptotic properties. 
In the following, we specify the internal and leaf tests.\sgap

{\bf Internal Test:} Suppose that the random walk arrives at a node on level $\level>0$.
A fixed number $\intsamples_{\level-1}$ of samples $\obs(i)$ is drawn from both children, and are used to compute the \acp{gllr}
\begin{equation}\label{eq:GLLR_Statistics}
\tilde{S}_{\text{GLLR}}^{(l-1)}(K_{l-1}) \triangleq \sum_{i=1}^{\intsamples_{\level-1}}\log
\frac{\density_{\level-1}\br{\obs(i)\gn\hat\paramvec_1^{(l-1)}}}
{\density_{\level-1}\br{\obs(i)\gn\paramvec_0^{(l-1)}}},
\end{equation}
 where $\hat\paramvec_1^{(l-1)}$ is the maximum likelihood estimate of the anomaly parameter, given by
\begin{equation}
\hat\paramvec_1^{(l-1)}=\argmax_{\paramvec \in\paramset_1^{(l-1)}}\prod_{i=1}^{\intsamples_{\level-1}}\density_{\level-1}\br{\obs(i)\gn\paramvec}.
\end{equation}
The statistics \eqref{eq:GLLR_Statistics} utilize the information on the normal distribution.
If at least one of the children has a strictly positive \ac{gllr}, the random walk moves to the child with the greater \ac{gllr}. Otherwise, it moves to the parent.
The sample size $K_\level$ for $\level=0,\dots,\levels-1$ is determined offline, such that the probability of moving in the direction of the anomaly is greater than $\frac{1}{2}$.
Note that $\intsamples_\level$ is finite under \ref{itm:distinguish}.
\vspace{0.1cm}

{\bf Leaf Test:}
When the random walk visits a leaf node, we perform an \ac{allr} test. Here, samples $\obs(i)$ are drawn sequentially from the process and the local \ac{allr} 
\begin{equation}
\tilde{S}_{\text{ALLR}}(\leaftime)=\sum_{i=1}^\leaftime\log
\frac{\density_0\br{\obs(i)\gn\hat\paramvec_1^{(0)}\br{i-1}}}{\density_0\br{\obs(i)\gn\paramvec_0^{(0)}}},
\label{eqn:ALLR}
\end{equation}
is continuously updated, where
\begin{equation}\label{eq:leaftestestimate}
\hat\paramvec_1^{(0)}(i-1)=\argmax_{\paramvec\in\paramset_1^{(0)}}\prod_{j=1}^{i-1}\density_0\br{\obs(j)\gn\paramvec},
\end{equation}
is the delayed maximum likelihood estimate of $\paramvec_1^{(0)}$.
To initialize the estimate $\hat\paramvec_1^{(0)}(0)$, a fixed number $\initleafn\geq0$ (which is independent of $\nproc,\cost$) of samples is drawn from the leaf.
In Appendix~\ref{app:proof1} we elaborate on how to set $\initleafn$.
As opposed to the \ac{gllr},   $\tilde{S}_{\text{ALLR}}(\leaftime)$ is a viable likelihood ratio, {so that the Wald likelihood ratio identity can still be applied to upper-bound the error probabilities of the  sequential test \cite{wald2004sequential}}.

At every time step $n>0$, the \ac{allr} \eqref{eqn:ALLR} is examined: if $\tilde{S}_{\text{ALLR}}(n)>\log\frac{\levels}{\cost}$, the random walk terminates and the tested process is declared anomalous, while a negative \ac{allr} results in returning to the parent node.
The resulting search policy is summarized in Algorithm~\ref{alg:AlgoHDS}.
%
 \begin{algorithm2e}
	\caption{Single Target HDS}
	\label{alg:AlgoHDS}
	\KwData{Inspected node at level $\level$}  
        \uIf{$l>0$ (internal node)}{
            Measure $\intsamples_{\level-1}$ samples from each child node\;
            Compute \ac{gllr} for each child via \eqref{eq:GLLR_Statistics}\; 
            \uIf{Both \acp{gllr} are negative}
                { 
                Invoke Algorithm~\ref{alg:AlgoHDS} on parent node\;
                }
            \uElse{
                Invoke Algorithm~\ref{alg:AlgoHDS} on child with larger \ac{gllr}\;
            }
        }
        \uElse{
            Init $\paramvec_1^{(0)}$ according to (\ref{eq:app_init}) and $n=1$\;
            \label{stp:ALLR} Draw $y(n)$ and compute \ac{allr}  \eqref{eqn:ALLR}\;
            \uIf{$\tilde{S}_{\text{ALLR}}(n)>\log\frac{\levels}{\cost}$}{
                Identify node as target and {\bf terminate}\;
                }
                \uElseIf{$\tilde{S}_{\text{ALLR}}(n) < 0$}
                {
                  Invoke Algorithm~\ref{alg:AlgoHDS} on parent node\;
                }
                Increment $n$ and jump  to step \ref{stp:ALLR}\;
        }
\end{algorithm2e}

\subsection{Performance Analysis}\label{subsec:analysis}
We next theoretically analyze the \ac{hds} policy, denoted $\hds$, for $K=1$. In particular, we establish that $\hds$ is asymptotically optimal in $\cost$, i.e.,
\begin{equation}
\lim_{\cost\rightarrow 0}\frac{\risk(\hds)}{\risk^*}=1,
\end{equation}
and order optimal in $\nproc$, namely,
\begin{equation}
\lim_{\nproc\rightarrow\infty}\frac{\risk(\hds)}{\risk^*}=\order(1)
\end{equation}
where $\risk^*$ is a lower bound on the Bayesian risk. This is stated in the following theorem:

\begin{theorem}
\label{th:Bayes}
When \ref{itm:distinguish}
holds and $\paramset_1^{(\level)}$ is finite for all $0\leq\level\leq\levels-1$, the Bayesian risk of $\hds$ is bounded by
\begin{equation}\label{eq:theorem1}
\risk(\hds)\leq
\cost\const\levels+
\frac{\cost\log\frac{\levels}{\cost}}
{\div_0\br{\paramvec_1^{(0)}||\paramvec_0^{(0)}}}+\order(\cost)\;,
\end{equation}
where $\const$ is a constant independent of $\nproc$ and $\cost$.
\end{theorem} 

\begin{IEEEproof}
\ifproofs
The complete proof 
is given in Appendix~\ref{app:proof1}.
\else 
Due to page limitations, the detailed proof can be found in \cite{Wolff2022hierarchical_dropbox}.
\fi
\begin{figure}[tbp]
\centerline{
\begin{tikzpicture}[
    every path/.append style={thick},
    scale=.5
    ]
    \def\off{1}
    \def\offs{1.3}
    \def\yoff{.7}
    \node[node2, 
    fill=white,
    ] (0) at (0,4.5){};
    \foreach\x [count=\xi] in {-4,4}{
        \node[node2, fill=white] (\x) at (\x,3){};
        \draw (\x) -- (0);
        \foreach\y [count=\yi] in {\x-2,\x+2}{
            \node[node2, fill=white] (\y) at (\y,1.5){};
            \draw (\x) -- (\y);
            \foreach\z [count=\zi] in {\y-1,\y+1}{
            \node[node2, fill=white] (\z) at (\z,0){};
            \draw (\z) -- (\y);
            \node at (\y+\off,0-1){};
        }
        }
    }
    \node[node2, fill=red] (anom) at (-3,0) {};
    \node (anomtext) at (-3,-1.4) [align=center] {$\uparrow$\\anomaly};
    \foreach\h [count=\l] in {0,1.5,3,4.5}{
        \pgfmathsetmacro\ldef{\l-1}
        \node (level) at (-9,\h) {$\level=\pgfmathprintnumber{\ldef}$};
    }
    \begin{pgfonlayer}{background}
        \node (t3) at (5,4.3){$\tree_3$};
        \draw[thick, fill=green!30]
        ([xshift=-.5cm]0.west) to[closed, curve through={
        ([xshift=-1cm]4-2.west)
        ([shift={(-.25cm,0)}]4-2-1.west)
        ([yshift=-1cm]4-2+1.south)
        ([shift={(+.2cm,-.2cm)}]4+2+1.east)
        ([shift={(+.0cm,+.5cm)}]4.north east)
        ([shift={(+.25cm,+.25cm)}]0.north east)
        }]  cycle;
        
        \node (t2) at (-6.2,3.5){$\tree_2$};
        \draw[thick, fill=blue!30]
        ([xshift=.5cm]-4.north east) to[closed, curve through={
        ([xshift=1.5cm]-4-2.east)
        ([shift={(.25cm,-.25cm)}]-4-2+1.south east)
        ([yshift=-2cm]-4-2.south)
        ([shift={(-.25cm,-.25cm)}]-4-2-1.south west)
        ([shift={(-.25cm,+.25cm)}]-4-2.north west)
        }]  cycle;
        
        \node (t1) at (-1,2){$\tree_1$};
        \draw[thick, fill=magenta!30]
        ([shift={(-.25cm,+.25cm)}]-4+2.north west) to[closed, curve through={
        ([shift={(+.25cm,+.25cm)}]-4+2+1.north)
        ([shift={(+.25cm,-.25cm)}]-4+2+1.south east)
        ([shift={(-1.4cm,+.25cm)}]-4+2+1.north)
        }]  cycle;
        
        \node (t0) at (-.9,-2){$\tree_0$};
        \draw[thick, fill=cyan!30]
        ([yshift=.25cm]-4+2-1.north) to[closed, curve through={
        ([shift={(.7cm,-1cm)}]-4+2-1.east)
        ([shift={(1cm,-2cm)}]-4+2-1.east)
        ([yshift=-2cm]-4+2-1.south)
        ([shift={(-1cm,-2cm)}]-4+2-1.west)
        ([shift={(-.7cm,-1cm)}]-4+2-1.west)
        }]  cycle;
    \end{pgfonlayer}
\end{tikzpicture}
}
\vspace{-0.2cm}
\caption{An illustration of the subtrees $\tree_0,\dots,\tree_\levels$ used in the analysis of the HDS algorithm.}
\vspace{-0.2cm}
\label{fig:subtrees}
\end{figure}
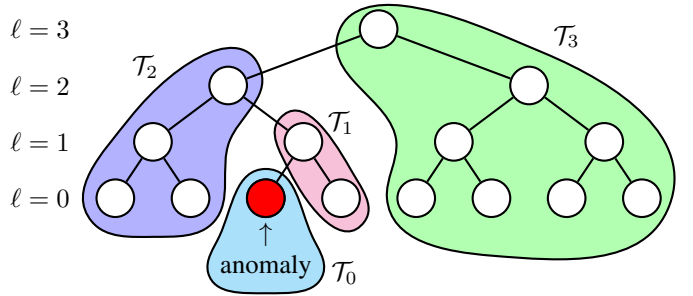
Here, we only present the proof outline,
which divides the trajectory of the random walk into two stages: {\em search} and {\em target test}.

In the \textit{search stage} the random walk explores the high level nodes and  is expected to eventually concentrate on the true anomaly.
Based on this insight, we partition the tree $\tree$ into a sequence of subtrees $\tree_0,\tree_1, \ldots,\tree_{\log_2M}$ (\Cref{fig:subtrees}). Subtree $\tree_{\log_2M}$ is obtained by removing the halftree that contains the target from $\tree$. Subtree $\tree_{\level}$ is iteratively obtained by removing the halftree that contains the target from $\tree\backslash\tree_{\level+1}$. $\tree_0$ consists of only the target node.
We then define the last passage time $\tau_{\level}$ of the search phase from each subtree $\tree_{\level}$. 
An upper bound on the end of this first stage is found by proving that the expected last passage time to each of the halftrees that do not contain the target is bounded by a constant. 
Summing the upper bound on the last passage times yields the first term in \eqref{eq:theorem1}.

The second stage is the {\em leaf target test}, which ends by declaring the target with expected time $\mathbb{E}[\tau_0]$.
To bound $\mathbb{E}[\tau_0]$, we first define a random time $\tau_{ML}$ to be the smallest integer such that the estimator of the target leaf's parameter equals to $\paramvec_1^{(0)}$ for all $n>\tau_{ML}$, and we show that $\mathbb{E}[\tau_{ML}]$ is bounded by a constant independent of $c$ and $M$. 
We then bound $\mathbb{E}[\tau_0]$ using Wald's equation \cite{wald2004sequential} and Lorden's inequality \cite{lorden1970excess}, which yields the second and third terms in \eqref{eq:theorem1}. 
Finally, we show that the detection error is of order $O(c)$. By using the martingale properties of the \ac{allr} statistic we prove that the false positive rate of the leaf test is bounded by $\frac{\cost}{\levels}$. In addition, the expected number of times a normal leaf is tested is in the order of $\levels$.  The resulting error rate $\er(\hds)$ is therefore in the order of $\cost$ (third term in \eqref{eq:theorem1}).
\end{IEEEproof}

The optimality properties of the Bayesian risk of \ac{hds} in both $c$ and $M$ directly carry through to the sample complexity of \ac{hds}, as stated in the following corollary:
\begin{corollary}\label{cor:sampleComplexity}
The sample complexity of $\mathrm{HDS}$ is bounded by:
\begin{align}
\complexity(\hds)&\leq \const\cdot\levels+\frac{\log\frac{\levels}{\cost}}
{\div_0\br{\paramvec_1^{(0)}||\paramvec_0^{(0)}}}+O(1)
\label{eqn:CompUp}\\
\complexity(\hds)&\geq\frac{\log_2 M}{I_{\text{max}}}+\frac{\log\frac{1-c}{c}}
{\div_0\br{\paramvec_1^{(0)}||\paramvec_0^{(0)}}}+O(1)
\label{eqn:CompDn}
\end{align}
where $I_{\text{max}}$ is the maximum mutual information between the true hypothesis and the observation under an optimal action. 
\end{corollary}

\begin{IEEEproof}
The upper bound \eqref{eqn:CompUp} follows directly from Theorem~\ref{th:Bayes}, while \eqref{eqn:CompDn} is obtained using \cite[Thm. 2]{naghshvar2013active}.
\end{IEEEproof}
Corollary \ref{cor:sampleComplexity} indicates that \ac{hds} is asymptotically optimal in $c$ and order optimal in $M$.

%
%
\subsection{Multi-Target Detection}\label{subsec:multi}
\begin{figure}[tbp]
\centerline{
\begin{tikzpicture}[
    every path/.append style={thick},
    scale=.5
    ]
    \def\off{1}
    \def\offs{1.3}
    \def\yoff{.7}
    \node[node2, fill=red] (anom) at (-3,0) {};
    \node[node2, fill=red] (anom2) at (-1,0) {};
    \node[node2, fill=red] (anom3) at (-2,1.5) {};
    \node[node2] (0) at (0,4.5){};
    \foreach\x [count=\xi] in {-4,4}{
        \node[node2] (\x) at (\x,3){};
        \draw (\x) -- (0);
        \foreach\y [count=\yi] in {\x-2,\x+2}{
            \node[node2] (\y) at (\y,1.5){};
            \draw (\x) -- (\y);
            \foreach\z [count=\zi] in {\y-1,\y+1}{
            \node[node2] (\z) at (\z,0){};
            \draw (\z) -- (\y);
            \node at (\y+\off,0-1){};
        }
        }
    }
    
    \node (box)[draw, densely dashed,
                  inner xsep=1ex, inner ysep=1ex,
                  fit=(anom)(anom2)(anom3)] {};
    \node (anomtext) at (-2,-2) [align=center,anchor=north] {declared\\anomalous};
    \draw (anomtext) [-stealth] edge ([yshift=-.9cm,xshift=.3cm]anom);
    \draw (anomtext) [-stealth] edge ([yshift=-.9cm,xshift=-.3cm]anom2);
    \foreach\h [count=\l] in {0,1.5,3,4.5}{
        \pgfmathsetmacro\ldef{\l-1}
        \node (level) at (-9,\h) {$\level=\pgfmathprintnumber{\ldef}$};
    }
\end{tikzpicture}
}
\vspace{-0.2cm}
\caption{Multi-target detection illustration. On the third run of the random walk, the nodes in the dashed box are no longer sampled from or visited.}
\vspace{-0.2cm}
\label{fig:binary_tree_multi_target}
\end{figure}
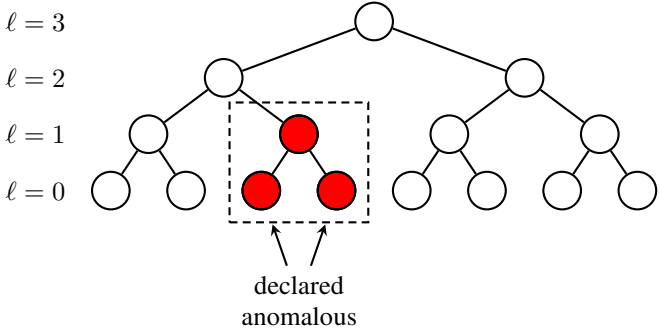

We next consider the detection of $\nanom>1$ anomalous processes. Our derivation and analysis is based on the following additional assumptions:

	\begin{enumerate}[label={\em AS\arabic*}, resume]   
	\item \label{itm:known} The number of anomalous processes $\nanom$ is a-priori known.
	\item \label{itm:remove} The search policy can remove a declared process from the tree, e.g., as in group testing the defective item is no longer tested in subsequent group tests.
	\item \label{itm:distinguish2} The distinguishability assumption \ref{itm:distinguish} is extended such that the distribution of a node that contains multiple anomalies is more similar to a node that contains a single anomaly, than to a normal node. To formulate mathematically, let $\paramset_j^{\br{\level}}$ be the set of parameters of a node that contains $j$ anomalies. We require, that there is $\mindelta>0$ such that \cref{eq:assumption} holds and that for all levels $\level=1,\dots,\levels$, number of anomalies $j=1,\dots,\min\br{\nanom,2^\level}$ and multi-anomaly parameter $\paramvec_j\in\paramset_j^{\br{\level}}$ it holds that 
	\begin{equation}\label{eq:assumption_multiple}
	\exists\paramvec_1^{(\level)}\in\paramset_1^{(\level)}: \div_\level\br{\paramvec_j||\paramvec_0^{(\level)}}-\div_\level\br{\paramvec_j||\paramvec_1^{(\level)}}\geq\mindelta.
	\end{equation}
	This assumption holds in a wide variety of scenarios and ensures that there is a bounded number of samples $\intsamples$ for the internal test, such that the random walk approaches the closest anomaly with a probability greater than $0.5$. 
\end{enumerate}

{\bf Algorithm Design:}
Since $\nanom$ is known by \ref{itm:known}, HDS formulated in Algorithm~\ref{alg:AlgoHDS} can be extended to locate the targets one-by-one. A process is declared anomalous by running the algorithm detailed in Subsection~\ref{subsec:HDSDesign}. This operation is feasible by \ref{itm:remove}. This means that subsequent random walks only visit nodes that contain undeclared processes (\Cref{fig:binary_tree_multi_target}). As a result, we only have to sample from one of the children during some internal tests. 

For the internal test, we still use the anomalous parameter sets $\paramset_1^{\br{\level}}$ that describe the distribution for \emph{one} anomaly within the node. This is justified by \ref{itm:distinguish2}. The resulting procedure is summarized as Algorithm~\ref{alg:AlgoHDS2}.

 \begin{algorithm2e}
	\caption{$\nanom$ Target HDS}
	\label{alg:AlgoHDS2}
	\KwData{Number of targets $\nanom$}  
	    \For{$k=1,\ldots,\nanom$}{
	        Identify $k$th target by invoking Algorithm~\ref{alg:AlgoHDS} at level $l=0$\;
	        Remove detected anomalous leaf node from tree\;
	    }
\end{algorithm2e}

{\bf Performance Analysis:} 
The theoretical guarantees derived for a single target in Subsection~\ref{subsec:analysis} carry also to the multi-target setting when \ref{itm:known}-\ref{itm:distinguish2} hold, in addition to \ref{itm:distinguish}. This is stated in the following theorem:

\begin{theorem}\label{th:Bayes_multiple}
When \ref{itm:distinguish}-\ref{itm:distinguish2}
hold, and $\paramset_1^{(\level)}$ is finite for all $0\leq\level\leq\levels-1$, the Bayesian risk of $\hds$ with $\nanom$ anomalous processes is bounded by:
\begin{equation}\label{eq:theorem2}
\risk(\hds)\leq
\cost\nanom\const\levels+
\frac{\cost\nanom\log\frac{\levels}{\cost}}
{\div_0\br{\paramvec_1^{(0)}||\paramvec_0^{(0)}}}+\order(\cost)\;,
\end{equation}
where $\const$ is a constant independent of $\nproc,\cost$ and $\nanom$.
\end{theorem} 

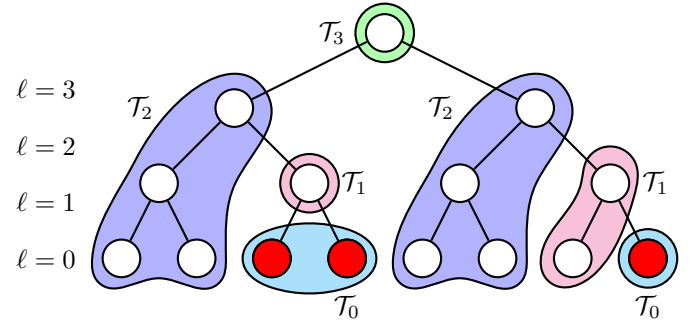
\begin{figure}
\centerline{
\begin{tikzpicture}[
    every path/.append style={thick},
    scale=.5
    ]
    \def\off{1}
    \def\offs{1.3}
    \def\yoff{.7}
    \node[node2, 
    fill=white,
    ] (0) at (0,6){};
    \foreach\x [count=\xi] in {-4,4}{
        \node[node2, fill=white] (\x) at (\x,4){};
        \draw (\x) -- (0);
        \foreach\y [count=\yi] in {\x-2,\x+2}{
            \node[node2, fill=white] (\y) at (\y,2){};
            \draw (\x) -- (\y);
            \foreach\z [count=\zi] in {\y-1,\y+1}{
            \node[node2, fill=white] (\z) at (\z,0){};
            \draw (\z) -- (\y);
        }
        }
    }
    \node[node2, fill=red] (anom) at (-3,0) {};
    \node[node2, fill=red] (anom) at (-1,0) {};
    \node[node2, fill=red] (anom) at (7,0) {};
    \foreach\h [count=\l] in {0,1.5,3,4.5}{
        \pgfmathsetmacro\ldef{\l-1}
        \node (level) at (-9,\h) {$\level=\pgfmathprintnumber{\ldef}$};
    }
    \begin{pgfonlayer}{background}
        \node (t3) at (-1.4,6){$\tree_3$};
        \draw[thick, fill=green!30]
        ([shift={(-.25cm,0cm)}]0.west) to[closed, curve through={
        ([shift={(0cm,+.25cm)}]0.north)
        ([shift={(+.25cm,0cm)}]0.east)
        ([shift={(0cm,-.25cm)}]0.south)
        }]  cycle;
        
        \node (t2) at (-6.5,4){$\tree_2$};
        \draw[thick, fill=blue!30]
        ([xshift=.5cm]-4.north east) to[closed, curve through={
        ([xshift=1.5cm]-4-2.east)
        ([shift={(.25cm,0cm)}]-4-2+1.east)
        ([shift={(.25cm,-.25cm)}]-4-2+1.south east)
        ([yshift=-2cm]-4-2.south)
        ([shift={(-.25cm,-.25cm)}]-4-2-1.south west)
        ([shift={(-.25cm,0cm)}]-4-2-1.west)
        ([shift={(-.25cm,+.25cm)}]-4-2.north west)
        }]  cycle;
        \node (t2) at (1.5,4){$\tree_2$};
        \draw[thick, fill=blue!30]
        ([xshift=.5cm]4.north east) to[closed, curve through={
        ([xshift=1.5cm]4-2.east)
        ([shift={(.25cm,0cm)}]4-2+1.east)
        ([shift={(.25cm,-.25cm)}]4-2+1.south east)
        ([yshift=-2cm]4-2.south)
        ([shift={(-.25cm,-.25cm)}]4-2-1.south west)
        ([shift={(-.25cm,0cm)}]4-2-1.west)
        ([shift={(-.25cm,+.25cm)}]4-2.north west)
        }]  cycle;
        
        \node (t1) at (-.8,2){$\tree_1$};
        \draw[thick, fill=magenta!30]
        ([shift={(-.25cm,0cm)}]-4+2.west) to[closed, curve through={
        ([shift={(0cm,+.25cm)}]-4+2.north)
        ([shift={(+.25cm,0cm)}]-4+2.east)
        ([shift={(0cm,-.25cm)}]-4+2.south)
        }]  cycle;
        \node (t1) at (7.2,2){$\tree_1$};
        \draw[thick, fill=magenta!30]
        ([shift={(-.25cm,.25cm)}]4+2.north west) to[closed, curve through={
        ([shift={(.25cm,.25cm)}]4+2.north east)
        ([shift={(.25cm,0cm)}]4+2.south east)
        ([shift={(.25cm,0cm)}]4+2-1.south east)
        ([shift={(-.25cm,-.25cm)}]4+2-1.south west)
        ([shift={(-.25cm,.25cm)}]4+2-1.north west)
        ([shift={(-.25cm,0cm)}]4+2.west)
        }]  cycle;
        
        \node (t0) at (-1,-1.3){$\tree_0$};
        \draw[thick, fill=cyan!30]
        ([yshift=.25cm]-4+2-1.north) to[closed, curve through={
        ([shift={(0cm,.25cm)}]-4+2+1.north)
        ([shift={(.25cm,0cm)}]-4+2+1.east)
        ([shift={(0cm,-.25cm)}]-4+2+1.south)
        ([yshift=-.25cm]-4+2-1.south)
        ([shift={(-.25cm,0cm)}]-4+2-1.west)
        }]  cycle;
        
        \node (t0) at (7,-1.3){$\tree_0$};
        \draw[thick, fill=cyan!30]
        ([shift={(-.25cm,0cm)}]4+2+1.west) to[closed, curve through={
        ([shift={(0cm,+.25cm)}]4+2+1.north)
        ([shift={(+.25cm,0cm)}]4+2+1.east)
        ([shift={(0cm,-.25cm)}]4+2+1.south)
        }]  cycle;
    \end{pgfonlayer}
\end{tikzpicture}
}
\vspace{-0.2cm}
\caption{Illustration of the tree partition $\tree_0,\dots,\tree_\levels$ used in the analysis of the HDS algorithm for multiple targets.}
\vspace{-0.2cm}
\label{fig:subtrees_multi}
\end{figure}

\begin{IEEEproof}
\ifproofs
The complete proof is given in \cref{app:proof_multiple}. Here, we only present the proof outline, which extends on the rationale of the proof of Theorem~\ref{th:Bayes}: Again, we 
divide the tree $\tree$ into a similar partition $\tree_0,\dots,\tree_\levels$, where the sets $\tree_\level$ are recursively obtained by removing the halftrees at level $\level$ that contain at least one anomaly from $\tree\setminus\tree_{\level+1}$ (\Cref{fig:subtrees_multi}). Roughly speaking, due to the assumption in \cref{eq:assumption_multiple}, the internal test and the leaf test have a greater probability of moving towards the \emph{closest} anomaly than away from it. This results in the same constant upper bound on the expected last passage times to the sets $\tree_1,\dots,\tree_\levels$ as in the single-target scenario, which implies that the first term in \cref{eq:theorem2} is the first term of \cref{eq:theorem1} scaled by the number of anomalies $\nanom$.
The leaf test is unaffected by the additional anomalies. Therefore, the sample complexity of a single random walk in the multi-target scenario has the same upper bound as in the single-target scenario resulting again in the sample complexity in the second and third terms being scaled by $\nanom$. Finally, the upper bound on the probability of the declaring a normal process anomalous remains unaffected too. Applying the union bound over the $\nanom$ random walks yields an error rate in the order of $\cost$ in the third term.
\end{IEEEproof}

\ifcorroltwo 
Similarly to risk guarantees, one can also bound the sample complexity of Algorithm~\ref{alg:AlgoHDS2}, as stated in the following:
\begin{corollary}\label{cor:sampleComplexity_mult}
The sample complexity of $\hds$ for the detection of $\nanom$ anomalies under \ref{itm:distinguish}-\ref{itm:distinguish2} is bounded via
\begin{align}
\complexity(\hds)&\leq\nanom\const\cdot\levels+\frac{\nanom\log\frac{\levels}{\cost}}
{\div_0\br{\paramvec_1^{(0)}||\paramvec_0^{(0)}}}+O(1).
\label{eqn:CompUp_mult}
\end{align}
\end{corollary}

\begin{IEEEproof}
The upper bound \eqref{eqn:CompUp_mult} follows directly from Theorem~\ref{th:Bayes_multiple}.
\end{IEEEproof}
Corollary \ref{cor:sampleComplexity_mult} and the lower bound  in \cref{eqn:CompDn} indicate that \ac{hds} is order optimal in $M$ and has an asymptotic ratio of $\nanom$ when $c$ approaches zero.
\fi
\color{black}
%

\subsection{Discussion}\label{subsec:discussion}
The proposed \ac{hds} algorithm is designed to efficiently search in hierarchical data structures while coping with an unknown anomaly distribution. It can be viewed as an extension of the \ac{irw} method \cite{wang2020information} to unknown anomaly parameters, while harnessing the existing knowledge regarding the distribution of the anomaly-free measurements.
The uncertainty in the anomaly distribution makes both the algorithm design and the performance analysis much more involved.
In contrast to existing hierarchical algorithms, \ac{hds} can incorporate general parameterized anomaly observation models, resulting in it being order optimal with respect to the search space size and asymptotically optimal in detection accuracy. 

The derivation of \ac{hds} motivates the exploration of several extensions. First, \ac{hds} is derived for hierarchical data that can be represented as a binary tree, while anomaly search with adaptive granularity may take the form of an arbitrary tree. 
In such case, the path length from each leaf to the root may be different, and thus the distribution of each node does not depend solely on its level on the tree. We conjecture that with some modifications on the HDS algorithm, optimal performances can be also guaranteed in this case. However, we leave this analysis for future work

Furthermore, we design \ac{hds} for detecting leaf targets, while in some scenarios one may have to cope with hierarchical targets, i.e., where intermediate nodes can be anomalous. 
An additional extension would be to consider a composite model for both normal and anomalous distributions.
Various models can be assumed in this case (i.e., identical/different parameter for all normal cells). Whether asymptotic optimality can be achieved under this setting remains open.
We leave the extension of \ac{hds} to these settings for future work.


%
\section{Numerical Evaluations}\label{sec:Simulations}
We next empirically compare HDS with the existing search strategies of \ac{ds} \cite{hemo2020searching}, \ac{irw} \cite{wang2020information}, and the \ac{cbrw} algorithm \cite{vakili2018hierarchical}.
The \ac{irw} algorithm has access to the true anomaly parameter $\paramvec_1^{(\level)}$, while the other algorithms only have access to $\paramset_1^{(\level)}$. 
\ac{irw} and \ac{hds} use fixed size internal tests that are not optimized for the specific simulation. Instead the sample sizes $\intsamples_\level$ are chosen as small as possible such that the desired drift towards the target is ensured. The performance of \ac{irw} should therefore be a best-case scenario for \ac{hds}. \ac{irw}, \ac{ds}, and \ac{hds} use $\cost=10^{-2}$, and \ac{cbrw} uses $p_0=0.2$ and $\epsilon=10^{-2}$. 
The values are averaged over $10^6$ Monte Carlo runs\footnote{The source code can be found in \url{https://github.com/DrummingBeb/Composite-Anomaly-Detection-via-Hierarchical-Dynamic-Search}.}.
%

\textbf{Scenario 1: Exponential Distributions}\\
We first simulate a scenario where the decision-maker observes the interoccurrence time of Poisson point processes with normal rate $\lambda_0=1$ and anomalous rate $\lambda_1=10^3$. The rates at the internal nodes are equal to the sum of the rates of their children.
The minimum rate that is considered anomalous is $\lambda_{1,\min}=\frac{\lambda_0+\lambda_1}{2}$ such that the anomaly parameter set is $\paramset_1^{(0)}=\closedleft{\lambda_{1,\min},\infty}$.
This scenario models the detection of heavy hitters among Poisson flows where the measurements are exponentially distributed packet inter-arrival times. \Ac{cbrw} uses the mean threshold $\statthresh_\level$, such that the generalized likelihood ratio is one at $\statthresh_\level$ and exact bounds for the mean of exponentially distributed random variables with rate $\lambda_\level=\frac{1}{\statthresh_\level}$. 

\begin{figure}
    \centerline{
    \begin{tikzpicture}
    \begin{axis}[
        results4,
        ymode=log,
        ]
        \addplot+[line, mark=square]  table{./figures/data/DS/bayes_risk_vs_levels_DS_EXP_k_1_fixed_size_False_lgllr_False_fork_2_c_0.01_lambda0_1_lambda1_1000.0_min_lambda1_500.5_sim_1000000.txt};
        \addlegendentryexpanded{DS}    
        \addplot+[line, mark=o]table{./figures/data/CBRW/bayes_risk_vs_levels_CBRW_EXP_k_1_fork_2_c_0.01_lambda0_1_lambda1_1000.0_min_lambda1_500.5_sim_9863902.txt};
        \addlegendentryexpanded{CBRW}        
        \addplot+[line, mark=triangle] table{./figures/data/HDS/lallr/fixed/bayes_risk_vs_levels_HDS_EXP_k_1_fixed_size_True_lgllr_False_fork_2_c_0.01_lambda0_1_lambda1_1000.0_min_lambda1_500.5_sim_1000000.txt};
        \addlegendentryexpanded{HDS}
        \addplot+[line, mark=star]
        table{./figures/data/IRW/bayes_risk_vs_levels_IRW_EXP_k_1_fork_2_c_0.01_lambda0_1_lambda1_1000.0_min_lambda1_1000.0_sim_7515183.txt};
        \addlegendentryexpanded{IRW
        }
    \end{axis}
    \end{tikzpicture}
    }
    \vspace{-0.2cm}
    \caption{Risk vs. number of processes, scenario 1.
    }
    \vspace{-0.2cm}
    \label{fig:sim1}
\end{figure}
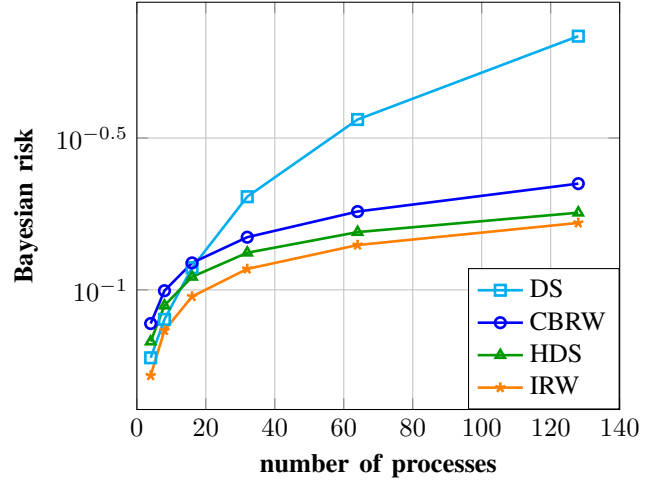

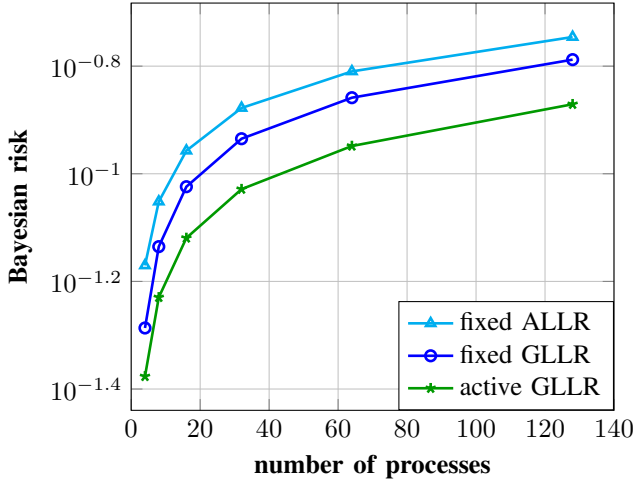
\begin{figure}
    \centerline{
    \begin{tikzpicture}
    \begin{axis}[
        results4,
        ymode=log,
        ]
        \addplot+[line, mark=triangle] table{./figures/data/HDS/lallr/fixed/bayes_risk_vs_levels_HDS_EXP_k_1_fixed_size_True_lgllr_False_fork_2_c_0.01_lambda0_1_lambda1_1000.0_min_lambda1_500.5_sim_1000000.txt};
        \addlegendentryexpanded{fixed ALLR}
        \addplot+[line, mark=o] table{./figures/data/HDS/lgllr/fixed/bayes_risk_vs_levels_HDS_EXP_k_1_fork_2_c_0.01_lambda0_1_lambda1_1000.0_min_lambda1_500.5_sim_1208647.txt};
        \addlegendentryexpanded{fixed GLLR}
        \addplot+[line, mark=star] table{./figures/data/HDS/lgllr/active/bayes_risk_vs_levels_HDS_EXP_k_1_fixed_size_False_lgllr_True_fork_2_c_0.01_lambda0_1_lambda1_1000.0_min_lambda1_500.5_sim_7930533.txt};
        \addlegendentryexpanded{active GLLR}
    \end{axis}
    \end{tikzpicture}
    }
    \vspace{-0.2cm}
\caption{\ac{hds} with different internal tests (fixed sample size vs. active) and leaf test statistics (\ac{allr} vs. \ac{gllr}), scenario 1. The active test uses a confidence level of $p=\frac{1}{2}+10^{-16}$.}
    \vspace{-0.2cm}
\label{fig:sim1active}
\end{figure}

\Cref{fig:sim1} depicts the risk $\risk\br{\policy}$ as in~\eqref{eqn:Risk} versus the number of processes $\nproc$. \textcolor{NewColor}{We can clearly observe that \ac{hds} outperforms \ac{cbrw} and \ac{ds} for most values, and it is within a minor gap of that of \ac{irw}.} \textcolor{NewColor}{While for $\nproc\geq16$, \ac{hds}  only slightly outperforms  \ac{cbrw}, it  notably outperforms \ac{ds}}. \textcolor{NewColor}{However,} it is noted that \ac{cbrw} uses sequential internal tests, which should be more efficient than the fixed size internal tests of \ac{hds}. For this reason, in this scenario we also compare an alternative internal test for \ac{hds}. The results of this study, depicted in \cref{fig:sim1active}, show that switching to the sequential  \ac{gllr} \textcolor{NewColor}{statistic} for the leaf test instead of the \ac{allr} \textcolor{NewColor}{statistic} yields a performance gain for all $\nproc$. An even greater jump in performance is achieved by using an active test for the internal nodes. The details of the active test are given in
\ifproofs
Appendix~\ref{app:active}.
\else
\cite[Appendix A]{Wolff2022hierarchical_dropbox}.
\fi

\textbf{Scenario 2: Bernoulli Interference}\\
Next, we simulate \textcolor{NewColor}{our decision making algorithm when considering a set of Poisson point processes} with rate $\lambda_0=0.1$. Here, the measurements of the nodes that contain the anomaly are corrupted by Bernoulli interference; i.e.,
\begin{equation}\label{eqn:IntBer}
\obs(i)\sim\Exp\br{2^\level\lambda_0}+z\cdot\sbr{-6+\br{a+6}\cdot\bernoulli\br{0.5}}.
\end{equation}
In \eqref{eqn:IntBer}, $z\in\{0,1\}$ indicates whether the node is anomalous, and $a$ is unknown.
The node parameter $\paramvec$ is given by the pair $\br{z,a}$, where $\paramvec_0^{(\level)}=\br{0,0}$,  $\paramvec_1^{(\level)}=\br{1,10}$, and $\paramset_1^{(\level)}=\set{1}\times\set{1,5,10}$ for all levels $0\leq\level\leq\levels$. \ac{cbrw} uses $\statthresh_\level=1$ and sub-Gaussian bounds with $\xi=0.05$.

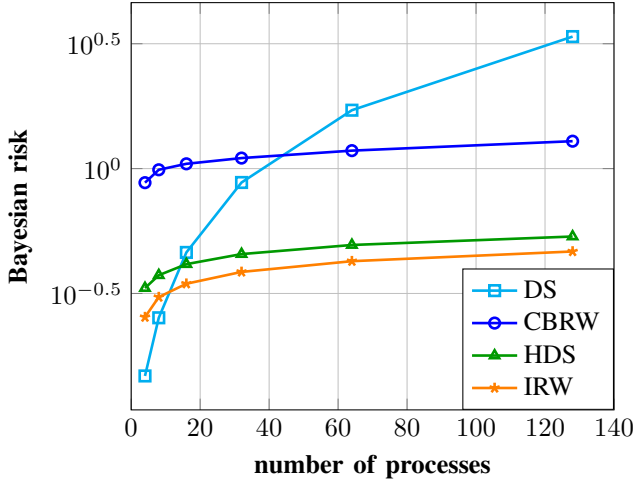
\begin{figure}
    \centerline{
    \begin{tikzpicture}
    \begin{axis}[
        results4,
        ymode=log,
        ]
        \addplot+[line, mark=square] table{./figures/data/DS/bayes_risk_vs_levels_DS_BERN_k_1_fixed_size_False_lgllr_False_fork_2_c_0.01_loc1_-6_lambda0_0.1_loc2_10.0_sim_1000000.txt};\addlegendentryexpanded{DS}
        \addplot+[line, mark=o]table{./figures/data/CBRW/bayes_risk_vs_levels_CBRW_BERN_k_1_fixed_size_False_lgllr_False_fork_2_c_0.01_loc1_-6_lambda0_0.1_loc2_10.0_sim_1000000.txt};\addlegendentryexpanded{CBRW}        
        \addplot+[line, mark=triangle] table{./figures/data/HDS/lallr/fixed/bayes_risk_vs_levels_HDS_BERN_k_1_fixed_size_True_lgllr_False_fork_2_c_0.01_loc1_-6_lambda0_0.1_loc2_10.0_sim_1000000.txt};
        \addlegendentryexpanded{HDS}
        \addplot+[line, mark=star]
        table{./figures/data/IRW/bayes_risk_vs_levels_IRW_BERN_k_1_fixed_size_True_lgllr_False_fork_2_c_0.01_loc1_-6_lambda0_0.1_loc2_10.0_sim_1000000.txt};
        \addlegendentryexpanded{IRW
        }
    \end{axis}
    \end{tikzpicture}
    }
    \vspace{-0.2cm}
    \caption{Risk vs. number of processes, scenario 2.}
    \vspace{-0.2cm}
    \label{fig:sim2}
\end{figure}

In this case the mean \textcolor{NewColor}{values} of the normal and abnormal distribution \textcolor{NewColor}{are} close to each other, and the anomalous \textcolor{NewColor}{process} {is reflected by higher moments of the distributions.} The results for this setting, depicted in \Cref{fig:sim2}, show that \textcolor{NewColor}{while \ac{cbrw} achieves poor performance,}
\Ac{hds} detects the anomaly efficiently, resulting in a larger gap between \ac{hds} and \ac{cbrw} than in the first scenario. 

\begin{figure}
\centering
\begin{tikzpicture}
    \begin{axis}[
        results_512,
        ymode=log,
        ]
        \addplot+[line, mark=square] table{./figures/data/multi_anomaly/bayes_risk_vs_levels_DS_EXP_k_5_fixed_size_False_lgllr_False_fork_2_c_0.01_lambda0_1_lambda1_1000.0_min_lambda1_500.5_sim_1000000.txt};
        \addlegendentryexpanded{DS}
        \addplot+[line, mark=o] table{./figures/data/multi_anomaly/bayes_risk_vs_levels_CBRW_EXP_k_5_fixed_size_False_lgllr_False_fork_2_c_0.01_lambda0_1_lambda1_1000.0_min_lambda1_500.5_sim_1000000.txt};
        \addlegendentryexpanded{CBRW}
        \addplot+[line, mark=triangle] table{./figures/data/multi_anomaly/bayes_risk_vs_levels_HDS_EXP_k_5_fixed_size_False_lgllr_True_fork_2_c_0.01_lambda0_1_lambda1_1000.0_min_lambda1_500.5_sim_1000000.txt};
        \addlegendentryexpanded{HDS}
        \addplot+[line, mark=star] table{./figures/data/multi_anomaly/bayes_risk_vs_levels_IRW_EXP_k_5_fixed_size_False_lgllr_False_fork_2_c_0.01_lambda0_1_lambda1_1000.0_min_lambda1_1000.0_sim_1000000.txt};
        \addlegendentryexpanded{IRW}
    \end{axis}
    \end{tikzpicture}
    \caption{Bayesian risk vs. number of processes, scenario 3.}
\label{fig:five_anomalies}
\end{figure}
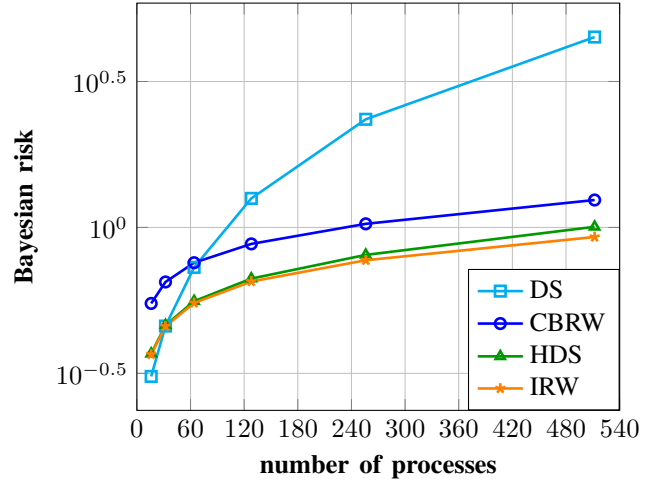

\textbf{Scenario 3: Multiple Anomalies}\\
Here we extend scenario 1 to $\nanom=5$ anomalies. \Ac{hds} and \ac{irw} use active internal tests. Additionally \Ac{hds} uses the \ac{gllr} statistic for the leaf tests. \Cref{fig:five_anomalies} shows a very similar picture as \Cref{fig:sim1}, in which \ac{hds} performs close to \ac{irw} and better than \ac{cbrw} and \ac{ds}. However, the performance \ac{hds} surpasses the non-hierarchical \ac{ds} at $\nproc=30$ processes as opposed to after already $\nproc=10$ processes in scenario 1.

\textbf{Scenario 4: Denial of Service Detection}\\
In this scenario, we detect \ac{dos} attacks using the DARPA intrusion detection data set \cite{dataset}.
Every entry in the data set corresponds to a packet arriving at an interface.
We only consider the timestamp, packet size and label (either normal or \ac{dos} traffic) of each packet.
The anomalous process ($\nanom=1$) corresponds to an interface that receives \ac{dos} traffic, so we simulate with permutations the entire data set. The normal processes are simulated by permutations of the packets that are labeled as normal traffic. 

We use the model in \cite{hemo2020searching} that considered a sample entropy for packet-size modeling, and demonstrated strong performance in detecting anomalous data on the DARPA data set.
Every 100ms seconds a sample is drawn by calculating the sample entropy of the packet sizes observed in the probed node during the current 100ms interval. Sampling from an internal node is naturally done by aggregating the packets of the processes within the node. The sample entropy is modeled with a Gaussian distribution that is parametrized by its mean and standard deviation. Using 1000 permutations of the training split ($50\%$ of the data), the distribution of the sample entropy is estimated for normal and anomalous nodes at all levels i.e. $\paramvec^{(\level)}_0=\big(\mu_0^{(\level)},\sigma^{(\level)}_0\big)$ and $\paramvec^{(\level)}_1=\big(\mu^{(\level)}_1,\sigma^{(\level)}_1\big)$ are estimated respectively for $\level=0,\dots,\levels-1$. The anomalous sample entropy is expected to have a smaller mean and variance i.e. $\mu_1^{(\level)}<\mu_0^{(\level)}$ and $\sigma^{(\level)}_1<\sigma^{(\level)}_0$. For \ac{ds} and \ac{hds},
the anomaly parameter sets are $\paramset_1^{(\level)}=\big(-\infty,\muhalf^{(\level)}\big]\times\big(0,\sigmahalf^{(\level)}\big]$ where $\muhalf^{(\level)}=\frac{\mu_0^{(\level)}+\mu_1^{(\level)}}{2}$ and $\sigmahalf^{(\level)}=\frac{\sigma_0^{(\level)}+\sigma_1^{(\level)}}{2}$. \Ac{hds} and \ac{irw} use active internal tests, and  \Ac{hds} uses the sequential \ac{gllr}  for the leaf tests. \Ac{cbrw} uses thresholds $\statthresh_\level=\frac{\mu_0^{(\level)}+\muhalf^{(\level)}}{2}$ and exact confidence intervals for the mean of normally distributed random variables with standard deviation $\sigma^{(\level)}=\frac{\sigma_0^{(\level)}+\sigmahalf^{(\level)}}{2}$. Due to instability of \ac{ds}, we discarded runs with more than 1000 samples. Therefore, the evaluation of \ac{ds} is very generous.

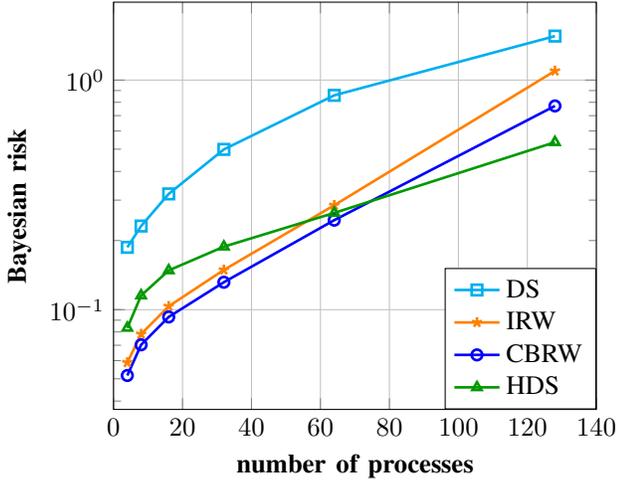
\begin{figure}
\centering
\begin{tikzpicture}
    \begin{axis}[
        results4,
        ymode=log,
        legend style={
        at={(1,0)},anchor=south east,cells={align=left}}
        ]
        \addplot+[line, mark=square] table{./figures/data/real_data/bayes_risk_vs_levels_DS_REAL_budget_None_p_nlelim_None_k_1_fixed_size_False_lgllr_False_max_samples_1000_fork_2_c_0.01_nothing_1.0_sim_1000000.txt};
        \addlegendentryexpanded{DS}
        \addplot+[line, mark=star, orange] table{./figures/data/real_data/bayes_risk_vs_levels_IRW_REAL_budget_None_p_nlelim_None_k_1_fixed_size_False_lgllr_False_fork_2_c_0.01_nothing_1.0_sim_20000.txt};
        \addlegendentryexpanded{IRW}
        \addplot+[line, mark=o, blue] table{./figures/data/real_data/bayes_risk_vs_levels_CBRW_REAL_budget_None_p_nlelim_None_k_1_fixed_size_False_lgllr_False_fork_2_c_0.01_nothing_1.0_sim_20000.txt};
        \addlegendentryexpanded{CBRW}
        \addplot+[line, mark=triangle, \mediumgreen] table{./figures/data/real_data/bayes_risk_vs_levels_HDS_REAL_budget_None_p_nlelim_None_k_1_fixed_size_False_lgllr_True_max_samples_-1_fork_2_c_0.01_nothing_1.0_sim_1000000.txt};
        \addlegendentryexpanded{HDS}
    \end{axis}
    \end{tikzpicture}
    \caption{Bayesian risk vs. number of processes, scenario 4.}
\label{fig:real_data}
\end{figure}

\Cref{fig:real_data} shows the risk as a function of the number processes. Interestingly, \ac{hds} scales better with the size of the search space when compared to the other hierarchical algorithms, namely \ac{irw} and \ac{cbrw}. We attribute this to the fact that the estimates can be inaccurate at high levels despite using a large training split and many permutations. \Ac{irw} loses performance because it relies on the point estimate $\paramvec_1^{(\level)}$ while the composite anomaly model of \ac{hds} is more robust.

\color{black}
\section{Conclusions}\label{sec:Conclusions}
In this work we developed a  sequential search strategy for the composite hierarchical anomaly detection problem dubbed \ac{hds}. \ac{hds} uses two variations of the \textcolor{NewColor}{\ac{gllr}} \textcolor{NewColor}{statistic} to ensure a biased random walk for a quick and accurate detection of the anomaly process. \Ac{hds} is shown to be order optimal with respect to the size of the search space and asymptotically optimal with respect to the detection accuracy. The addition of the hierarchical search significantly improves the performance over linear search methods in the common case of a large number of processes and heavy hitting anomalies.
We empirically show that the  performance can be further improved by using different statistics and local tests, and that for real-world data the composite anomaly model of \ac{hds} is more robust to inaccurate estimates from training  than existing algorithms that assume a known anomalous distribution model.
\ifproofs
\appendices
\crefalias{section}{appendix}
\section{Active Internal Test}\label{app:active}
Instead of the fixed size internal test described in \cref{subsec:HDSDesign}, we can use an active internal test:\\
Let $S_L(\time)$ and $S_R(\time)$ be the \ac{gllr} of the left and right children respectively at time $\time$ and initialize them with zero at $\time=0$.
As in the IRW active test \cite{wang2020information}, we define the thresholds
\begin{equation}
    v_0\triangleq-\log\frac{2p}{1-p},\quad v_1\triangleq\log\frac{2p}{1-p}
\end{equation}
where $p>\frac{1}{2}$ is the confidence level.
Let child
\begin{equation}
    x(\time-1)=\argmax_{i\in\set{\text{L},\text{R}}}S_i(\time-1)
\end{equation}
be the child with the higher \ac{gllr} at time $\time-1$.
Then, in every step $t$, we draw a sample from child $x(\time-1)$ and update $S_{x(t)}(\time)$. The other child $\tilde x(t)\neq x(\time)$ keeps the previous \ac{gllr} i.e., $S_{\tilde x(\time)}(t)=S_{\tilde x(\time)}(\time-1)$.
The test terminates at the random time
\begin{equation}
    k=\inf\set{\time\in\nat\gn S_{x(t)}(\time)\leq v_0\text{ or }S_{x(t)}(\time)\geq v_1}.
\end{equation}
If $S_{x(k)}(k)\geq v_1$, the random walk zooms into child $x(k)$ and if $S_{x(k)}(k)\leq v_0$, the random walk zooms out to the parent.

We observe a significant gain in empirical performance when compared to the fixed sample internal test (\Cref{fig:sim1active}).

\section{Proof of Theorem \ref{th:Bayes}}\label{app:proof1}
To find an upper bound on the Bayesian risk of \ac{hds}, we analyze the case where it is \textit{implemented indefinitely}, meaning that \ac{hds} probes the processes indefinitely according to its selection rule, while the stopping rule is disregarded.
We divide the trajectory of indefinite HDS into discrete steps at times $\time\in\nat$.
A step is not necessarily associated with every sample as will become clear later.
Let $\stop_\infty$ mark the first time that indefinite HDS performs a leaf test on the true anomaly and $\tilde{S}_{\text{ALLR}}$ rises above the threshold. It is easy to see that regular HDS terminates no later than $\stop_\infty$.
We divide the initial trajectory $\time=1,2,\dots,\stop_\infty$ of the indefinite random walk into two stages:
\begin{itemize}
    \item In the \textit{search stage} the random walk explores the high level nodes and eventually concentrates at the true anomaly. This stage ends at time $\stop_s$ which is the last time a leaf test is started on the true anomaly before $\stop_\infty$.
    \item The second stage is the \textit{target test} which ends with the declaration of the target. The duration of this stage is 
    $\stop_0$.
\end{itemize}

\textbf{Step 1: Bound the sample complexity of the search stage:}\\
We partition the tree $\tree$ into a sequence of sub-trees $\tree_0,\tree_1, \ldots,\tree_{\log_2M}$ (\Cref{fig:subtrees}) and define the last passage time $\stop_\level$ as described in \cref{subsec:analysis}. 
Let $\treeind(\time)$ indicate the sub-tree of the node tested at time $\time$.
The last passage time to $\tree_\levels$ is
\begin{equation}\label{eq:last_passage_time_root}
    \stop_\levels=\sup\set{\time\in\nat:\treeind(\time)=\tree_\levels}
\end{equation}
For the smaller sub-trees $\tree_1,\dots,\tree_{\levels-1}$ the last passage times are defined recursively such that
\begin{equation}\label{eq:last_passage_time}
    \stop_i=\sup\set{\time\in\nat:\treeind(\time)=\tree_i}-\stop_{i+1}.
\end{equation}
Notice, that the search time is bounded by
\begin{equation}\label{eq:searchstepsup}
    \stop_s=\sup_{1\leq\level\leq\levels}\stop_\level\leq\sum_{\level=1}^\levels\stop_\level.
\end{equation}
Next, we bound the expected last passage times $\expect[\stop_\level]$ for $1\leq\level\leq\levels$.
Towards this end, we define a distance $\dist_\time$ from the state of the indefinite random walk at time $\time$ to the anomalous leaf. When an internal node is probed, $\dist_\time$ is equal to the discrete distance to the anomaly on the tree. Since the walk starts at the root, we have $\dist_0=\levels$. when testing a normal leaf, $\dist_\time$ is equal to the sum of the discrete distance on the tree and the accumulated $\tilde{S}_{\text{ALLR}}$ of the current leaf test. When the true anomaly is probed, the distance is negative i.e. $\dist_\time=-\tilde{S}_{\text{ALLR}}$. Let the step $\step_\time$ be the random change in the distance at time $\time$ such that $\dist_{\time+1}=\dist_\time+\step_\time$.
Internal tests comprise only a single step either towards or away from the anomaly, i.e., $\step_\time\in\set{-1,1}$.
Because the sample sizes $\intsamples_\level$ of the internal tests are constructed such that $\mass\br{\step_\time=1}<\frac{1}{2}$, we have
\begin{equation}\label{eq:negexpect1}
    \expect\sbr{\step_\time
    }=2\mass\sbr{\step_\time=1}-1<0.
\end{equation}
We now show that if the sets of anomalous parameters $\paramset_1^{(\level)}$ are finite, there exists a bounded number of samples $\intsamples_\level$ such that \cref{eq:negexpect1} holds for the internal test at all levels.
We identify the two events
\begin{align}
    \eve_0&=\text{ the tested node does not contain the anomaly}\label{eq:two_events1}\\
    \eve_1&=\text{ the tested node contains the anomaly}.\label{eq:two_events2}
\end{align}
The probability of making a step in the wrong direction with an internal test is upper bounded by
\begin{align}\label{eq:internal_test_bound_error}
    \mass\sbr{\step_\time=1}\leq\max\br{\mass\sbr{\step_\time=1\gn\eve_0},\mass\sbr{\step_\time=1\gn\eve_1}}.
\end{align}
We first bound the first term in the maximization of \cref{eq:internal_test_bound_error}.
Let $\mass_{\paramvec_i}$ be the probability measure when the true state of nature is $\paramvec_i, i=0,1$, and let $\expect_{\paramvec_i}$ be the operator of expectation with respect to the measure $\mass_{\paramvec_i}$. 
Let $S_{\paramvec_0}$ and $S_{\paramvec_1}$ be the random \acp{gllr} based on $\intsamples$ samples from a normal node and an anomalous node respectively, where we omit the level $\level$ for readability. Then, under $\expect_{\paramvec_0}$ an error implies that at least one of the \acp{gllr} is strictly positive. By applying the union bound we get
\begin{equation}\label{eq:exponential_decay1}
    \mass\sbr{\step_\time=1\gn\eve_0}\leq2\mass\sbr{S_{\paramvec_0}>0}.
\end{equation}
Let 
$\tilde\paramvec=\argmax_{\paramvec\in\paramset}\prod_{i=1}^{\intsamples}\density\br{\obs(i)\gn\paramvec} $
be the \ac{mle} in the set $\paramset=\set{\paramvec_0}\cup\paramset_1$.
The event that $S_{\paramvec_0}$ is strictly positive implies that $\tilde\paramvec\neq\paramvec_0$ via the definition of the \ac{mle}.
Therefore, we find that
\begin{align}
    \mass\sbr{S_{\paramvec_0}>0}=\sum_{\paramvec_1\in\paramset_1}&\mass_{\paramvec_0}\sbr{\tilde\paramvec=\paramvec_1}\label{eq:exponential_decay2}.
\end{align}
Applying the definition of the \ac{mle}, the Chernoff bound and the \ac{iid} property yields
\begin{align}
    \mass_{\paramvec_0}\sbr{\tilde\paramvec=\paramvec_1}&\leq\mass_{\paramvec_0}\sbr{\sum_{i=1}^\intsamples\log\frac{\density\br{\obs(i)\gn\paramvec_1}}{\density\br{\obs(i)\gn\paramvec_0}}\geq0}\nonumber\\
    &\leq\br{\expect_{\paramvec_0}\sbr{\exp\br{-s\log\frac{\density\br{\obs(i)\gn\paramvec_0}}{\density\br{\obs(i)\gn\paramvec_1}}}}}^\intsamples\label{eq:this_decays_exponentially}
\end{align}
for all $s\geq0$. Notice, that the derivative of the expectation on the RHS of \cref{eq:this_decays_exponentially} with respect to $s$,
   $ -\div\br{\paramvec_0||\paramvec_1}\leq-\mindelta<0,$
is strictly negative for all $\paramvec_1$ due to the assumption in \cref{eq:assumption}. Thus, for all $\paramvec_1\in\paramset_1$ there exists a $s>0$ such that the RHS of \cref{eq:this_decays_exponentially} decays exponentially meaning that there exist a bounded $C>0$ and a $\gamma>0$ such that
\begin{align}\label{eq:exponential_decay3}
    \mass_{\paramvec_0}\sbr{\tilde\paramvec=\paramvec_1}\leq Ce^{-\gamma\intsamples}.
\end{align}
Combining \cref{eq:exponential_decay1}, \cref{eq:exponential_decay2} and \cref{eq:exponential_decay3}, we find that $\mass\sbr{\step_\time=1|\eve_0}$ decays exponentially with the number of samples $\intsamples$.

Next, we show that $\mass\sbr{\step_\time=1|\eve_1}$ also decays exponentially. Under $\eve_1$, the event that the \ac{gllr} of the anomalous child is strictly positive and the \ac{gllr} of the normal child is negative implies, that we move towards the anomaly, resulting in
\begin{align}
    &\mass\sbr{\step_\time=1\gn\eve_1}=1-\mass\sbr{\step_\time=-1\gn\eve_1}\nonumber\\ \nonumber
    &\leq1-\mass\sbr{S_{\paramvec_1}>0}\cdot\mass\sbr{S_{\paramvec_0}\leq0}
    \leq\mass\sbr{S_{\paramvec_1}\leq0}+\mass\sbr{S_{\paramvec_0}>0}.
\end{align}
We already showed that $\mass\sbr{S_{\paramvec_0}>0}$ decays exponentially with $\intsamples$, it remains to show the same for $\mass\sbr{S_{\paramvec_1}\leq0}$. Using the definition of the \ac{mle}, the Chernoff bound and the \ac{iid} property find
\begin{align}
    \mass\sbr{S_{\paramvec_1}\leq0}&\leq\mass_{\paramvec_1}\sbr{\sum_{i=1}^\intsamples\log\frac{\density(\obs(i)\gn\hat\paramvec_1)}{\density\br{\obs(i)\gn\paramvec_0}}\leq0}\nonumber\\
    &\leq\mass_{\paramvec_1}\sbr{\sum_{i=1}^\intsamples\log\frac{\density\br{\obs(i)\gn\paramvec_1}}{\density\br{\obs(i)\gn\paramvec_0}}\leq0}\nonumber\\
    &\leq\br{\expect_{\paramvec_1}\sbr{\exp\br{-s\log\frac{\density\br{\obs(i)\gn\paramvec_1}}{\density\br{\obs(i)\gn\paramvec_0}}}}}^\intsamples.\label{eq:this_decays_exponentially2} 
\end{align}
for all $s\geq0$.
Once again, the derivative of the expectation on the RHS of \cref{eq:this_decays_exponentially2} with respect to $s$,
   $ -\div\br{\paramvec_1||\paramvec_0}\leq-\mindelta<0, $
is strictly negative for all $\paramvec_1$ due to the assumption in \cref{eq:assumption}.
It follows that $\mass\sbr{\step_\time=1\gn\eve_1}$ decays exponentially with the number of samples $\intsamples$. Thus, there exists a bounded $\intsamples$ such that \cref{eq:negexpect1} holds.

On leaf nodes, every single sample of the sequential test comprises a step. A step is therefore the change in $\tilde{S}_{\text{ALLR}}$. Using the assumption in (\ref{eq:assumption}) and the independence of $\hat\paramvec_1^{(0)}(i-1)$ and $\obs(i)$ we find that for \emph{normal} leafs
\begin{equation}\label{eq:negexpect2}
    \expect\sbr{\step_\time}=\expect_{\paramvec_0^{(0)}}\sbr{\log\frac{\density_0\br{\obs(\time)\gn\hat\paramvec_1^{(0)}(\time-1)}}{\density_0\br{\obs(\time)\gn\paramvec_0^{(0)}}}}\leq-\mindelta<0.
\end{equation}
Similarly, we want to show that for the \emph{anomalous} leaf that
\begin{equation}\label{eq:negexpect3}
    \expect\sbr{\step_\time}=\expect_{\paramvec_1^{(0)}}\sbr{-\log\frac{\density_0\br{\obs(\time)\gn\hat\paramvec_1^{(0)}(\time-1)}}{\density_0\br{\obs(\time)\gn\paramvec_0^{(0)}}}}<0.
\end{equation}
Denoting $\hat\paramvec=\hat\paramvec_1^{(0)}(\time-1)$, we split the term and use the law of total expectation to find that
\begin{align}
    &\nonumber\expect\sbr{\step_\time}=
    \expect_{\paramvec_1^{(0)}}\sbr{-\log\frac{\density_0\br{\obs(\time)\gn\hat\paramvec}}{\density_0\br{\obs(\time)\gn\paramvec_0^{(0)}}}+\smash{\underbrace{\log\frac{\density_0\br{\obs(\time)\gn\paramvec_1^{(0)}}}{\density_0\br{\obs(\time)\gn\paramvec_1^{(0)}}}}_{=0}}}\\[12pt] 
    &=-\div_0\br{\paramvec_1^{(0)}\,||\,\paramvec_0^{(0)}}+\mass_{\paramvec_1^{(0)}}\sbr{\hat\paramvec\neq\paramvec_1^{(0)}}\div_0\br{\paramvec_1^{(0)}\,||\,\hat\paramvec}
\end{align}
where we used the fact that $\div_0\br{\paramvec_1^{(0)}\,||\,\paramvec_1^{(0)}}=0$.
For \cref{eq:negexpect3} to hold, it remains to be shown that
\begin{equation} 
\label{eq:boundedprob}
    \mass_{\paramvec_1^{(0)}}\sbr{\hat\paramvec\neq\paramvec_1^{(0)}} 
     <\inf_{\hat\paramvec\in\paramset_1^{0}}\frac{\div_0\br{\paramvec_1^{(0)}\,||\,\paramvec_0^{(0)}}}{\div_0\br{\paramvec_1^{(0)}\,||\,\hat\paramvec}}\triangleq\lambda_{\paramvec_1^{(0)}}.
\end{equation}
Notice, that the $\lambda_{\paramvec_1^{(0)}}$ are strictly positive due to the assumption in \cref{eq:assumption} and assuming that
    $\sup_{\paramvec_1^{(0)},\hat\paramvec\in\paramset_1^{0}}\div_0\br{\paramvec_1^{(0)}\,||\,\hat\paramvec}<\infty.$
For this purpose, we first introduce the following Lemma: 

\begin{lemma}
Let $\paramset_1^{(0)}$ be finite, i.e., $\setsize=|\paramset_1^{(0)}|<\infty$ and let $\hat\paramvec_1^{(0)}(\leaftime)$ be the ML estimate of $\paramvec_1^{(0)}$ using $\leaftime$ samples. Let $\tau_{ML}$ be the smallest integer such that $\hat\paramvec_1^{(0)}(\leaftime)=\paramvec_1^{(0)}$ for all $\leaftime>\tau_{ML}$.
Then, there exist a bounded $C>0$ and a $\gamma>0$ independent of $\nproc$ and $\cost$ such that
\begin{equation}
    \mass_{\paramvec_1^{(0)}}\sbr{\tau_{ML}>\leaftime}\leq Ce^{-\gamma\leaftime}.
\end{equation}
\end{lemma}
\begin{IEEEproof}
    The event $\tau_{ML}>\leaftime$ implies that there exists a time $\time>\leaftime$ such that $\hat\paramvec_1^{(0)}(\time)\neq\paramvec_1^{(0)}$ and therefore we have
\begin{equation}\label{eq:dsproof}
    \mass_{\paramvec_1^{(0)}}\sbr{\tau_{ML}>\leaftime}\leq\sum_{\time=\leaftime}^\infty\mass_{\paramvec_1^{(0)}}\sbr{\hat\paramvec_1^{(0)}(\time)\neq\paramvec_1^{(0)}}.
\end{equation}
By definition of the maximum likelihood estimate, the event $\hat\paramvec_1^{(0)}(\time)\neq\paramvec_1^{(0)}$ implies
    $\sum_{i=1}^\time S_{\tilde\paramvec}(i)\geq 0 $
for some $\tilde\paramvec\neq\paramvec_1^{(0)}$, where
 $   S_{\tilde\paramvec}(i)=\log\frac{\density\br{\obs(i)|\tilde\paramvec}}{\density\br{\obs(i)|\paramvec_1^{(0)}}}.$
Applying the Chernoff bound and using the \acs{iid} property yields
\begin{equation}\label{eq:decay}
    \mass_{\paramvec_1^{(0)}} \sbr{\sum_{i=1}^\time S_{\tilde\paramvec}(i)\geq0}\leq\br{\expect_{\paramvec_1^{(0)}}\sbr{e^{sS_{\tilde\paramvec}(i)}}}^\time
\end{equation}
for all $s\geq0$. The moment generating function (MGF) $e^{sS_{\tilde\paramvec}(i)}$ is equal to one at $s=0$. The derivative of the MGF at $s=0$ is
$    \expect_{\paramvec_1^{(0)}}\sbr{S_{\tilde\paramvec}(i)}=-\div_0\br{\paramvec_1^{(0)}||\tilde\paramvec}<0.$
Because the derivative is negative and assuming that the distribution of $S_{\tilde\paramvec}(i)$ is light-tailed\footnote{A distribution with density $\density$ is light-tailed if $\int_{-\infty}^\infty e^{\lambda x}\density(x)dx<\infty$ for some $\lambda>0$ \cite{lighttailed}.}, there exist $s>0$ and $\gamma>0$ such that $\expect\sbr{e^{sS_{\tilde\paramvec}(i)}}=e^{-\gamma}<1$ and the RHS of (\ref{eq:decay}) decays exponentially with $\time$. Summing over all $\tilde\paramvec\neq\paramvec_1^{(0)}$, we get
$\mass_{\paramvec_1^{(0)}}\sbr{\hat\paramvec_1^{(0)}(\time)\neq\paramvec_1^{(0)}}\leq\setsize e^{-\gamma\time},$
and thus the RHS of (\ref{eq:dsproof}) is bounded by
$    \sum_{\time=\leaftime}^\infty\setsize e^{-\gamma\time}=\frac{\setsize}{1-e^{-\gamma}}e^{-\gamma n}.$
\end{IEEEproof}

In light of lemma 1, we propose the following mechanism to ensure that \cref{eq:boundedprob} holds: Whenever a leaf test is started, before beginning with the sequential test described in \cref{subsec:HDSDesign}, a fixed number $\initleafn\geq0$ of samples $\set{\obs_i}_{i=-\initleafn+1}^0$ is drawn from the leaf to initialize the estimate $\hat\paramvec_1^{(0)}$, meaning, instead of \cref{eq:leaftestestimate} we write
\begin{equation}
    \label{eq:app_init}
    \hat\paramvec_1^{(0)}(i-1)=\argmax_{\paramvec\in\paramset_1^{(0)}}\prod_{j=-\initleafn+1}^{i-1}\density_0\br{\obs(j)\gn\paramvec}.
\end{equation}
This has the effect, that at every step of the subsequent sequential test, the estimate $\hat\paramvec_1^{(0)}$ is based on at least $\initleafn$ samples.
Since $\hat\paramvec\neq\paramvec_1^{(0)}$ implies that $\stop_{ML}>\initleafn$, we have
\begin{equation}
    \mass_{\paramvec_1^{(0)}}\sbr{\hat\paramvec\neq\paramvec_1^{(0)}}\leq\mass_{\paramvec_1^{(0)}}\sbr{\stop_{ML}>\initleafn}.
\end{equation}
Using 
 $   \lambda=\inf_{\paramvec_1^{(0)}\in\paramset_1^{(0)}}\lambda_{\paramvec_1^{(0)}}$
and lemma 1 we find that \cref{eq:boundedprob} is satisfied if
   $ \initleafn>-\frac{\log\frac{\lambda}{C}}{\gamma}.$
Notice, that $\initleafn$ is chosen independent of the size of search space $\nproc$ and the cost $\cost$.

With (\ref{eq:negexpect1}), (\ref{eq:negexpect2}) and (\ref{eq:negexpect3}) we established that HDS has the same drift behavior as IRW. Furthermore, we assume that the distribution of
    $\log\frac{\density_0\br{\obs(i)\gn\tilde\paramvec}}{\density_0\br{\obs(i)\gn\paramvec_0^{(0)}}}$
is light-tailed for all $\tilde\paramvec\in\paramset_1^{(0)}$.

Thus, we can apply \cite[Lemma 1,2]{wang2020information} and find that the expected last passage times $\expect[\tau_i]$ for $1\leq i\leq\levels$ are bounded by a constant $\passtimebound$ independent of $\nproc$ and $\cost$. Applying (\ref{eq:searchstepsup}) yields
\begin{equation}\label{eq:searchstepbound}
    \expect[\stop_s]\leq\passtimebound\levels.
\end{equation}
Let
 $   \maxintsamples=\sup_{0\leq\level\leq\levels-1} \{ \intsamples_\level\}$
be the maximum number of samples taken from a child during an internal test. Then every step $\step_\time$ takes at most $\maxsamples=\max\set{2\maxintsamples,\initleafn+1}$ samples and the complexity of the search stage $\complexity_s$ is bounded by
\begin{equation}\label{eq:searchcomplexity}
    \complexity_s\leq\maxsamples\expect[\stop_s]\leq\const\log_2\nproc
\end{equation}
where $\const=\passtimebound\maxsamples$ is a constant independent of $\nproc$ and $\cost$.

\textbf{Step 2: Bound the sample complexity of the target test:}\\
In the analysis of the target test we associate a time step $\leaftime=1,2,\dots,\stop_0$ with every sample. Using lemma 1 and the tail sum for expectation we find 
\begin{equation}\label{eq:estimate}
    \expect\sbr{\tau_{ML}}=\order(1).
\end{equation}
At all times $\leaftime>\tau_{ML}$, we necessarily have $\hat\paramvec_1^{(0)}=\paramvec_1^{(0)}$. From the definition of $\tilde S_\text{LALLR}$ in (\ref{eqn:ALLR}) it is easy to see, that after $\leaftime=\tau_{ML}+1$, the leaf test is essentially a sequential likelihood ratio test. The expected time until the threshold $\log\frac{\levels}{c}$ is reached $\stop_f=\stop_0-\stop_{ML}$ is bounded by
\begin{equation}\label{eq:terminate}
    \expect[\stop_f]\leq\frac{\log\frac{\levels}{c}}{\div_0\br{\paramvec_1^{(0)}||\paramvec_0^{(0)}}}+\order(1)
\end{equation}
where we used Wald's equation \cite{wald2004sequential} and Lorden's inequality \cite{lorden1970excess} and assumed that the first two moments of the log-likelihood ratio are finite. Combining (\ref{eq:estimate}) and (\ref{eq:terminate}) yields the sample complexity of the target test
\begin{equation}\label{eq:leafcomplexity}
    \complexity_t=\expect\sbr{\tau_0}\leq\frac{\log\frac{\levels}{c}}{\div_0\br{\paramvec_1^{(0)}||\paramvec_0^{(0)}}}+\order(1).
\end{equation}

\textbf{Step 3: Bound the error rate:}\\
Notice, that detection errors can only occur in the search stage. The expected number of times a normal leaf is tested $\expect[N]$ is bounded by the number of steps in the search stage. Thus, using (\ref{eq:searchstepbound}) we get
\begin{equation}\label{eq:normalleaftest}
    \expect[N]\leq\expect[\stop_s]\leq\passtimebound\levels. 
\end{equation}
Let $Z(\leaftime)=\exp\br{\tilde{S}_{\text{ALLR}}(\leaftime)}$ be adaptive likelihood ratio at time $\leaftime$.
In the following, we use the properties of the ALLR to bound the false positive rate of the leaf test
\begin{equation}
    \alpha=\mass_{\paramvec_0^{(0)}}\sbr{Z(\leaftime)\geq\frac{\levels}{\cost}\text{ for some }\leaftime\geq1}.
\end{equation}
Note that on normal leafs $Z(n)$ is a non-negative martingale, i.e.,
\begin{align}
    &\expect_{\paramvec_0^{(0)}}\sbr{Z(n+1)\gn\set{\obs(i)}_{i=1}^n}\\
    &=Z(n)\expect_{\paramvec_0^{(0)}}\sbr{\frac{\density\br{\obs(n+1)\gn\hat\paramvec_1^{(0)}(n)}}{\density\br{\obs(n+1)\gn\paramvec_0^{(0)}}}}
    =Z(n)
\end{align}
where we used the independence of $\hat\paramvec_1^{(0)}(n)$ and $\obs(n+1)$ in the last step.
Using a lemma for nonnegative supermartingales \cite{martingale} we find
\begin{equation} \nonumber
    \mass_{\paramvec_0^{(0)}}\sbr{Z(\leaftime)\geq\frac{\levels}{\cost}\text{ for some }\leaftime\geq1}\\
    \leq\frac{\cost}{\levels}\expect_{\paramvec_0^{(0)}}\sbr{Z(1)}.
\end{equation}
Since
 $   Z(1)=\expect_{\paramvec_0^{(0)}}\sbr{\frac{\density\br{\obs(1)\gn\hat\paramvec_1^{(0)}(0)}}{\density\br{\obs(1)\gn\paramvec_0^{(0)}}}}=1,$
the false positive rate is bounded by
\begin{equation}\label{eq:falsepositiverate}
    \alpha\leq\frac{\cost}{\levels}.
\end{equation}
Finally, combining (\ref{eq:normalleaftest}) and (\ref{eq:falsepositiverate}) yields the bound on the error rate
\begin{equation}\label{eq:errorrate}
    \er(\hds)\leq\alpha\cdot\expect[N]\leq
    \passtimebound\cost=\order(\cost)
\end{equation}

Theorem 1 follows from (\ref{eq:searchcomplexity}), (\ref{eq:leafcomplexity}) and (\ref{eq:errorrate}).

\section{Proof of \Cref{th:Bayes_multiple}}\label{app:proof_multiple}
To find an upper bound on the Bayesian risk of \ac{hds} in the multi-target scenario, we analyze the $\nanom$ random walks separately. This can be done because there is at least one undeclared anomalous leaf in the tree $\tree$ during each random walk.\\
\textbf{Step 1: Bound the sample complexity of the search stage:}\\
Similar to the proof in \cref{app:proof1}, we divide the tree $\tree$ as described in \cref{subsec:multi} and \cref{fig:subtrees_multi}. The last passage times are defined recursively by \crefrange{eq:last_passage_time_root}{eq:last_passage_time} and the search time is bounded by \cref{eq:searchstepsup}. Let $\dist^{(i)}_t$ be the distance to the $i$-th anomalous leaf at time $t$, where the distance is defined as in \cref{app:proof1}. Now consider the change in the distance to the \emph{closest} anomaly $\step_t=\dist_{t+1}-\dist_t$
where
   $ \dist_t=\min_i\dist^{(i)}_t. $
We want to show that in expectation the minimum distance decreases at all times during the random walk i.e. 
\begin{equation}\label{eq:negexpect_general}
\expect\sbr{\step_t}<0.    
\end{equation}
As the leaf test is unaffected by additional anomalies and the currently tested leaf is also the closest, it only remains to show that \cref{eq:negexpect_general} holds for the internal test. Recall, that the number of samples $\intsamples_\level$ of an internal test is chosen such that \cref{eq:negexpect_general} holds. In \cref{app:proof1}, we have proven that such a $\intsamples_\level$ exists for the two events $\eve_0$ and $\eve_1$ defined in \crefrange{eq:two_events1}{eq:two_events2}. Notice, that under $\eve_0$ the closest anomaly lies outside the tested node and the distance to it is in expectation reduced by moving to the parent by following the same argument as for a single anomaly. 
Now, we recognize the events
\begin{equation}
    \eve_j=\text{ the tested node contains $j$ anomalies}
\end{equation}
for $j\geq1$. Notice, that the $j$ anomalies within the node are the closest anomalies and they are equally close. Moving to a child that contains at least one anomaly reduces $\dist_t$ by 1. We distinguish the two events
\begin{align}
    \eve_j^{(1)}&=\text{ one of the children contains anomalies}\\
    \eve_j^{(2)}&=\text{ both of the children contains anomalies}.
\end{align}
Let $S_{\paramvec_j}$ be the random \acp{gllr} based on $\intsamples'$ samples from a node containing $j$ anomalies, where we omit the level $\level$ for readability. Then under $\eve_j^{(1)}$, the event that the \ac{gllr} of the anomalous child is strictly positive and the \ac{gllr} of the normal child is negative, implies $\step_t=-1$ such that
\begin{align}
    & \mass\sbr{\step_\time=1\gn\eve_j^{(1)}}=1-\mass\sbr{\step_\time=-1\gn\eve_j^{(1)}}\nonumber\\
    &\leq1-\mass\sbr{S_{\paramvec_j}>0}\cdot\mass\sbr{S_{\paramvec_0}\leq0} \nonumber 
    \leq\mass\sbr{S_{\paramvec_j}\leq0}+\mass\sbr{S_{\paramvec_0}>0}.
\end{align}
We already showed that $\mass\sbr{S_{\paramvec_0}>0}$ and $\mass\sbr{S_{\paramvec_1}\leq0}$ decay exponentially with $\intsamples'$ (\Cref{app:proof1}), it remains to show the same for $\mass\sbr{S_{\paramvec_j}\leq0}$ with $j>1$. Using the definition of the \ac{mle}, the Chernoff bound and the \ac{iid} property find
\begin{align}
    \mass\sbr{S_{\paramvec_j}\leq0}&\leq\mass_{\paramvec_j}\sbr{\sum_{i=1}^{\intsamples'}\log\frac{\density(\obs(i)\gn\hat\paramvec_1)}{\density\br{\obs(i)\gn\paramvec_0}}\leq0}\nonumber\\
    &\leq\mass_{\paramvec_j}\sbr{\sum_{i=1}^{\intsamples'}\log\frac{\density\br{\obs(i)\gn\paramvec_1}}{\density\br{\obs(i)\gn\paramvec_0}}\leq0}\nonumber\\
    &\leq\br{\expect_{\paramvec_j}\sbr{\exp\br{-s\log\frac{\density\br{\obs(i)\gn\paramvec_1}}{\density\br{\obs(i)\gn\paramvec_0}}}}}^{\intsamples'}.\label{eq:this_decays_exponentially_multi} 
\end{align}
for all $\paramvec_1\in\paramset_1$ and $s\geq0$. Due to the assumption in \cref{eq:assumption_multiple}, for all $\paramvec_j\in\paramset_j$ there exists a $\paramvec_1$ such that the derivative of the expectation on the RHS of \cref{eq:this_decays_exponentially_multi} with respect to $s$
\begin{equation}
    \div_\level\br{\paramvec_j||\paramvec_1}-\div_\level\br{\paramvec_j||\paramvec_0}\leq-\mindelta<0.
\end{equation}
is strictly negative. Therefore $\mass\sbr{S_{\paramvec_j}\leq0}$ and $\mass\sbr{\step_\time=1\gn\eve_j^{(1)}}$ decay exponentially with $\intsamples'$.

Next, we consider $\eve_j^{(2)}$. Moving away from the closest anomalies implies that the \ac{gllr} of both children is negative such that
\begin{align}\label{eq:final_multi_bound}
    \mass\sbr{\step_\time=1\gn\eve_j^{(2)}}=\mass\sbr{S_{\paramvec_{j_l}}\leq0}\cdot\mass\sbr{S_{\paramvec_{j_r}}\leq0}.
\end{align}
where $\paramvec_{j_l}$ and $\paramvec_{j_r}$ are the parameters of the left and right child containing $j_l$ and $j_r$ anomalies respectively. The factors on the RHS of \cref{eq:final_multi_bound} decay exponentially with $\intsamples'$.
It follows that there exists a bounded number of samples $\intsamples'$ such that \cref{eq:negexpect_general} holds.


Following the same arguments as in step 1 of \cref{app:proof1}, we find that the sample complexity of a single random walk is bounded by \cref{eq:searchcomplexity}. Consequently, the complexity of the search stages of the $\nanom$ random walks is bounded by
\begin{equation}\label{eq:searchcomplexity_multi}
    \complexity_s\leq\nanom\const\log_2\nproc.
\end{equation}
\textbf{Step 2: Bound the sample complexity of the target test:}
Since, the leaf target test is unaffected by additional anomalies, its sample complexity is bounded by \cref{eq:leafcomplexity} and summing over the $\nanom$ random walks yields
\begin{equation}\label{eq:leafcomplexity_multi}
    \complexity_t\leq\nanom\expect\sbr{\tau_0}\leq\frac{\nanom\log\frac{\levels}{c}}{\div_0\br{\paramvec_1^{(0)}||\paramvec_0^{(0)}}}+\order(1).
\end{equation}
\textbf{Step 3: Bound the error rate:} 
Applying the reasoning in step 3 of \cref{app:proof1} we find that the error rate is bounded by \cref{eq:errorrate} and applying the union bound over the $\nanom$ random walks yields
\begin{equation}\label{eq:errorrate_multi}
    \er(\hds)=\nanom\alpha\expect[N]\leq
    \nanom\passtimebound\cost=\order(\cost).
\end{equation}

\Cref{th:Bayes_multiple} follows from \cref{eq:searchcomplexity_multi}, \cref{eq:leafcomplexity_multi} and \cref{eq:errorrate_multi}.
\color{black}

\fi

\bibliographystyle{IEEEtran}
\bibliography{IEEEabrv,bib}

\end{document}